\documentclass[journal=inoraj]{achemso}
\setkeys{acs}{usetitle = true}
\usepackage{rotating,lineno}
\usepackage[version=3]{mhchem} 
\RequirePackage{url}

\SectionNumbersOn

\author{Sinhu\'e L\'opez-Moreno$^1$}
\affiliation[Universidad Aut\'onoma de Campeche]
{$^1$CONACYT - Centro de Investigaci\'on en Corrosi\'on, Universidad Aut\'onoma de Campeche, Av. 
H\'eroe de Nacozari 480, Campeche, Campeche 24029, M\'exico}
\email{sinlopez@uacam.mx}

\author{Pl\'acida Rodr\'iguez-Hern\'andez$^3$}
\affiliation[Malta and Universidad de La Laguna]
{$^3$MALTA Consolider Team, Departamento de F\'isica, Instituto de Materiales y 
Nanotecnolog\'ia, and Malta Consolider Team, Universidad de La Laguna, La Laguna 
38205, Tenerife}

\author{Alfonso Mu\~noz$^3$}
\affiliation[Malta and Universidad de La Laguna]
{$^3$MALTA Consolider Team, Departamento de F\'isica, Instituto de Materiales y 
Nanotecnolog\'ia, and Malta Consolider Team, Universidad de La Laguna, La Laguna 
38205, Tenerife}

\author{Daniel Errandonea$^2$}
\affiliation[MALTA and Universitad de Valencia]
{$^2$MALTA Consolider Team, Departamento de F\'isica Aplicada-ICMUV, Universitad de 
Valencia, Edificio de Investigaci\'on, c/Dr. Moliner 50, Burjassot, 46100 Valencia, 
Spain}

\title{First-principles study of InVO$_4$ under pressure: phase transitions from CrVO$_4$- to AgMnO$_4$-type structure}


\begin{document}

\setlength{\fboxrule}{0 pt}
\begin{tocentry}
\centering
\includegraphics[width=8.5cm]{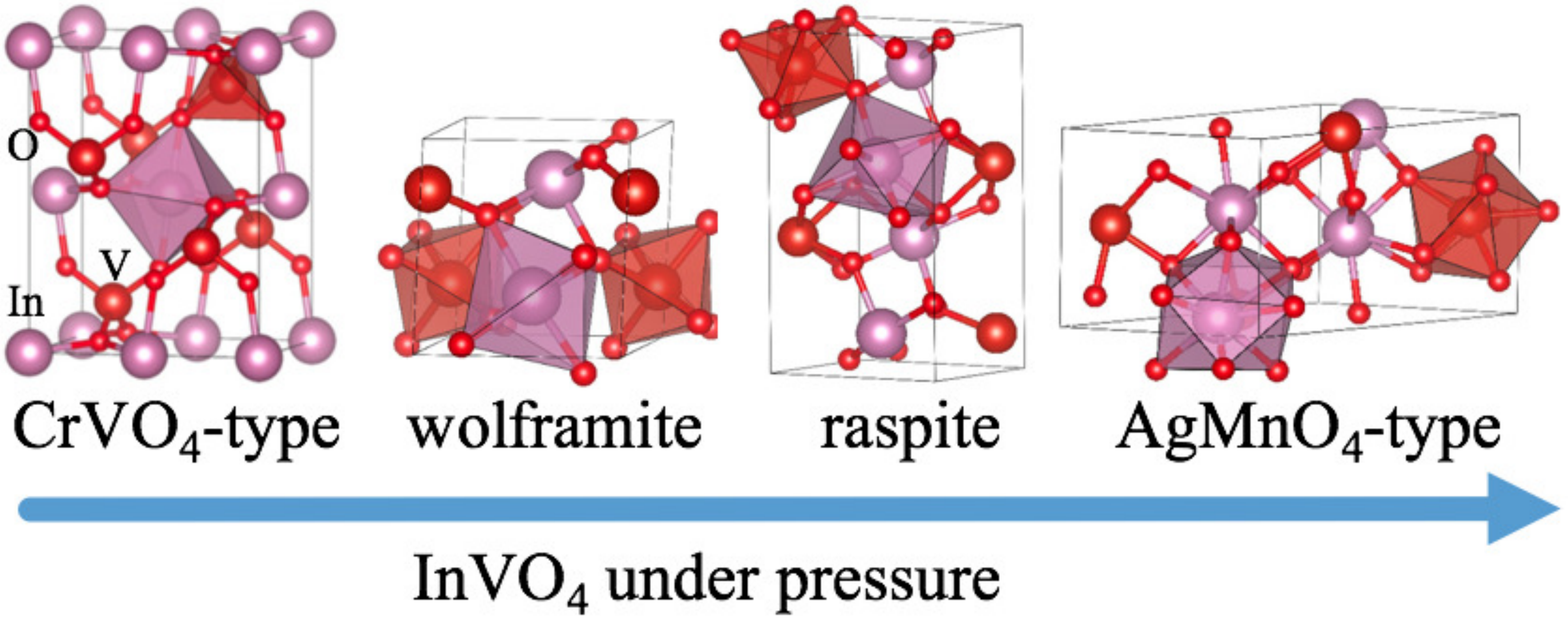} 
The high-pressure stability of InVO$_4$ is investigated using \textit{ab initio} calculations. 
Where the phase transitions driven by pressure were determined within the quasiharmonic 
approximation at 300 K. In our findings two new HP post-wolframite phases were obtained. 
The equation of state, crystal structure, phonon spectrum and electronic structure were 
studied for each stable phase as well as their evolution with pressure.
\\
{\bf {\Large For Table of Contents Only}}
\end{tocentry}

\pagebreak

\begin{abstract}
First-principles calculations have been done to study the InVO$_4$ compound under 
pressure. In this work, total energy calculations were performed in order to analyze the structural behavior of the experimentally known polymorphs of InVO$_4$: 
$\alpha$-MnMoO$_4$-type (I), CrVO$_4$-type (III), and the wolframite (V). Besides, 
in this paper we present our results about the stability of this compound beyond the 
pressures reached by experiments. We propose some new high pressure phases based in 
the study of 13 possible candidates. The quasiharmonic approximation has been used to
calculate the sequence of phase transitions at 300 K: CrVO$_4$-type, III (in parentheses 
the transition pressure) $\rightarrow$ wolframite, V (4.4 GPa) $\rightarrow$ raspite, VI 
(28.1 GPa) $\rightarrow$ AgMnO$_4$-type, VII (44 GPa). Equations of state and phonon
frequencies as function of pressure have been calculated for the studied phases. In order 
to determine the stability of each phase we also report the phonon dispersion along the
Brillouin zone and the phonon density of states for the most stable polymorphs. Finally, 
the electronic band structure for the low- and high-pressure phases for the studied 
polymorphs is presented as well as the pressure evolution of the band gap by using 
the HSE06 hybrid functional.  
\end{abstract}

\section{Introduction}
Vanadates $A$VO$_4$~\cite{Errandonea2008} oxides have been the focus of many 
studies due to their wide physical properties which leads to important applications in 
various fields, such as thermophosphorus sensors, high-power lasers, scintillators, 
active material for gas sensors, catalysis for water splitting and electrolyte for lithium 
ion batteries, to name a 
few.~\cite{Errandonea2008,Rapaport1999,Rapaport1999b,Lempicki1993,Allison1995, 
Baran1998,Ai2010,Lin2007,Butcher2010,Zou2001} Several vanadates can crystallize 
in structures such as zircon [space group (SG): $I4_1/amd$, No. 141, $Z$= 4, crystal
structure (CS): tetragonal], ~\cite{Panchal2011,Errandonea2009} scheelite (SG: 
$I4_1/a$, No. 88, $Z$ = 4, CS: tegragonal),~\cite{Tomeno2011} and monazite (SG: 
$P2_1/n$, No. 14, $Z$ = 4, CS: monoclinic),~\cite{Clavier2011} however, to our 
knowledge, only the vanadates CrVO$_4$-III, FeVO$_4$-II, TlVO$_4$ and 
InVO$_4$, have been synthesized in the CrVO$_4$-type structure (SG: $Cmcm$, No. 
63, $Z$ = 4, CS: orthorhombic).~\cite{Baran1998} While CrVO$_4$-III and 
FeVO$_4$-II have been studied by using several 
techniques,~\cite{Touboul1995b,Tojo2006,Oka1996,Robertson1972,Muller1975} 
only the thermodynamic properties,~\cite{Touboul1980} the vibrational 
spectra,~\cite{Baran1985} and the photoelectrochemical response~\cite{Butcher2010} 
were reported for TlVO$_4$. In contrast, there have been more studies dedicated to 
study InVO$_4$ due to its potential for applications such as a catalyst for production 
of hydrogen by visible-light driven water 
splitting.~\cite{Ai2010,Enache2009,Lin2007,Krol2011,Butcher2010,Zou2001} 

\begin{figure}[htb]
\centering
\begin{tabular}{cccccc}
\includegraphics[width=2.8cm]{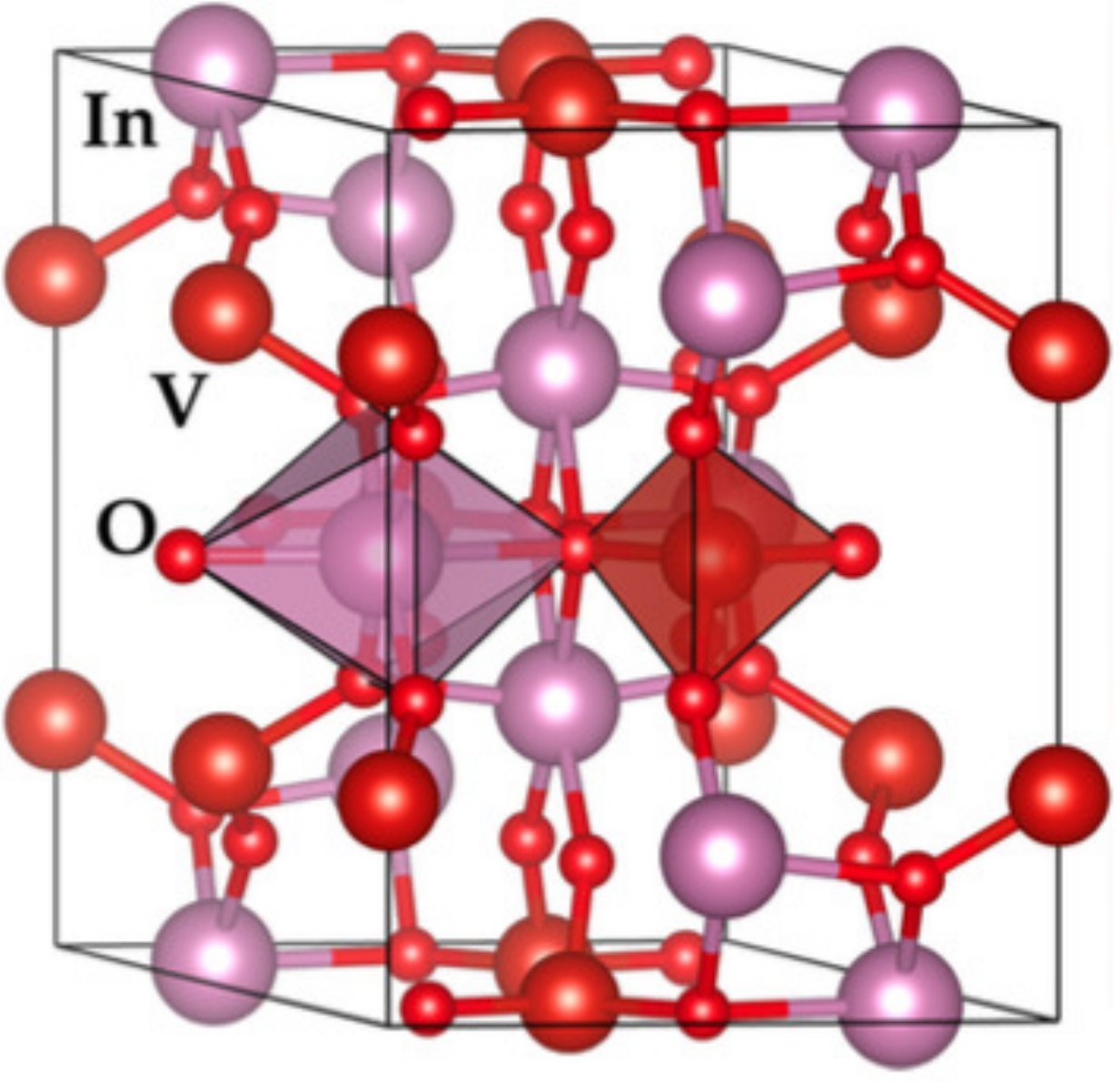} & 
\includegraphics[width=2.2cm]{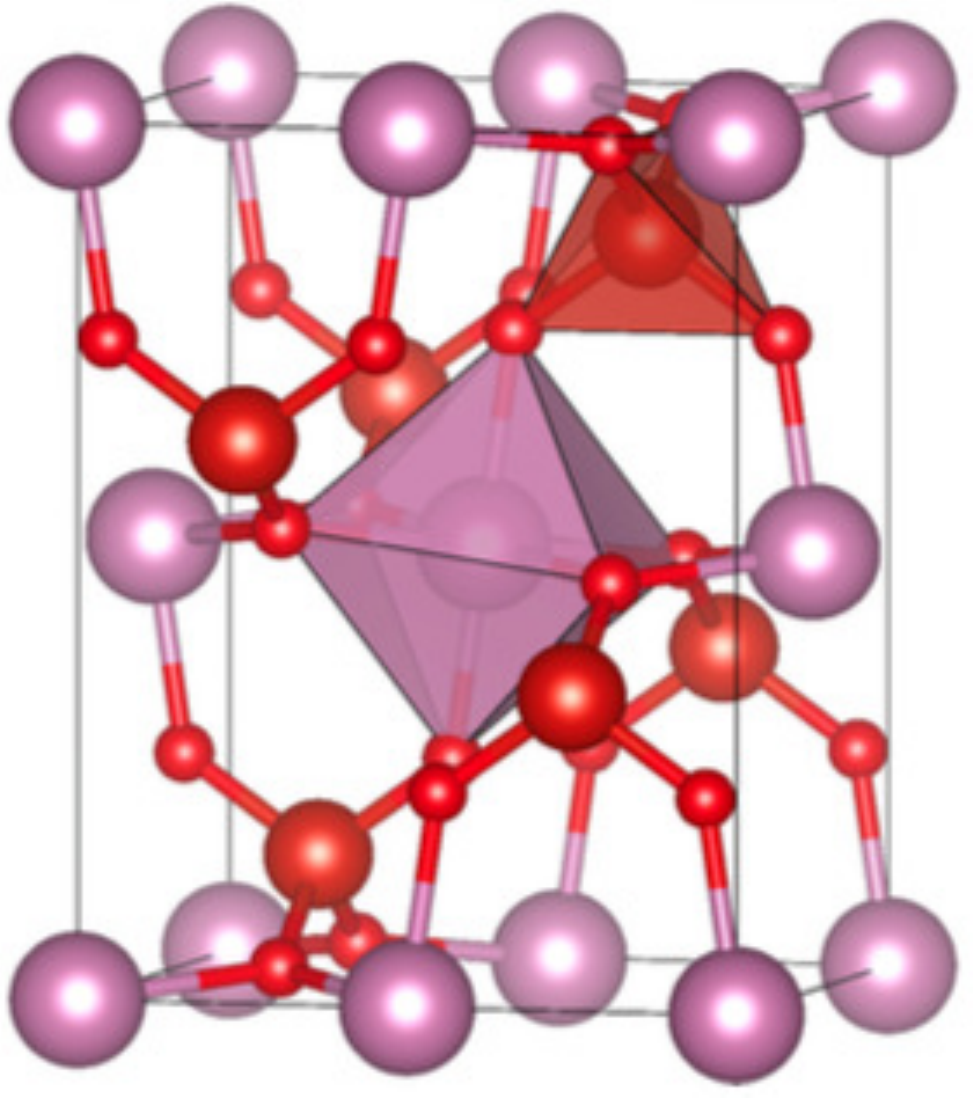}  &
\includegraphics[width=2.1cm]{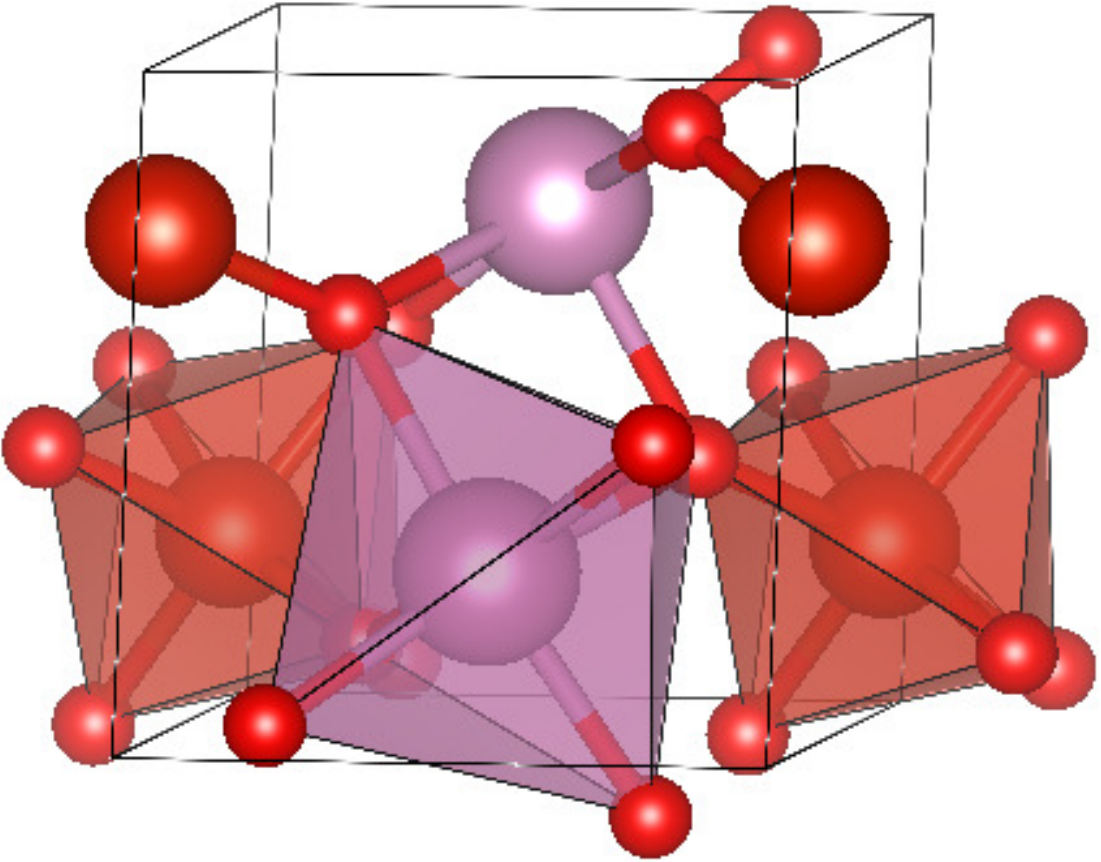} &
\includegraphics[width=2.1cm]{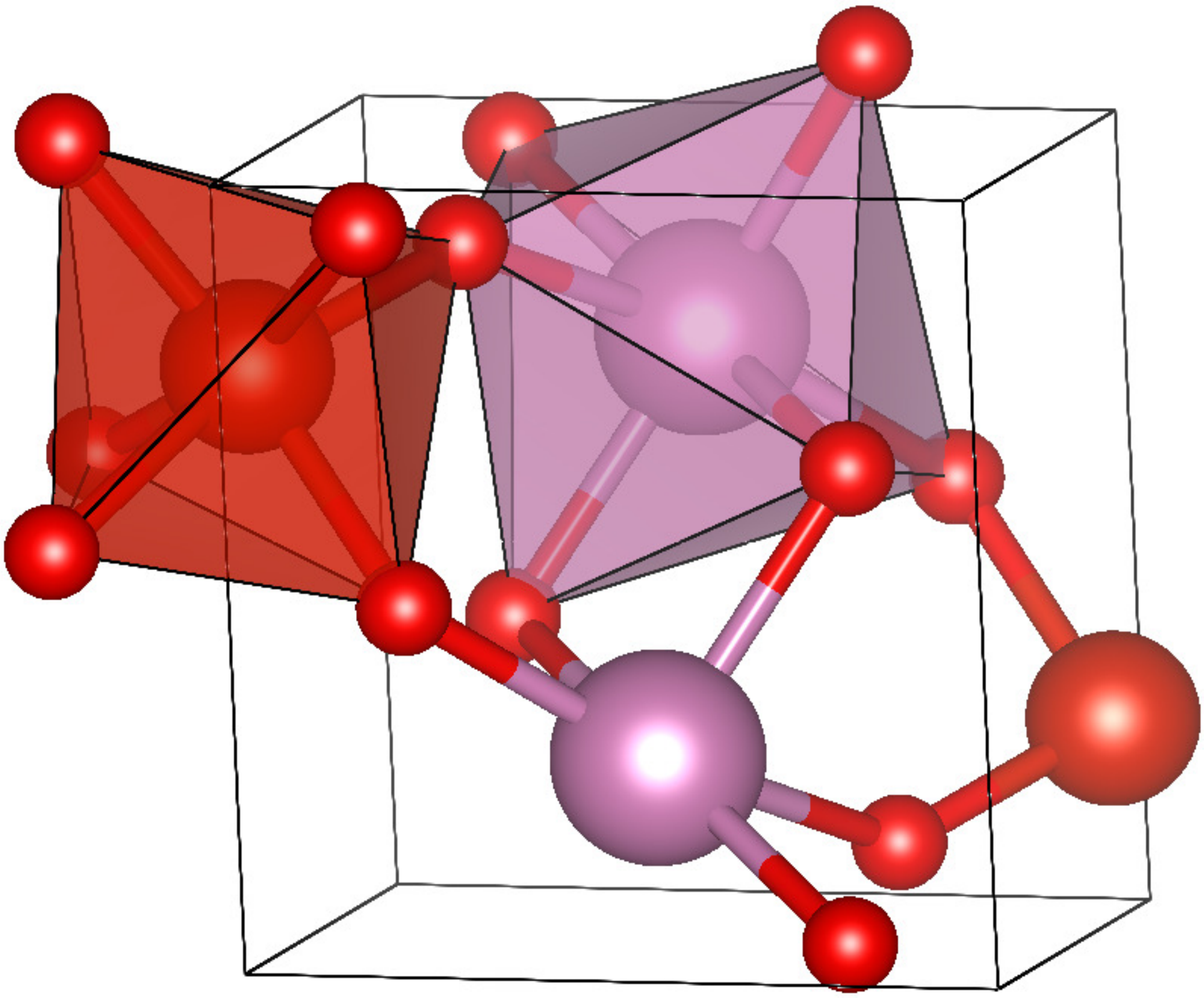}  &
\includegraphics[width=2.4cm]{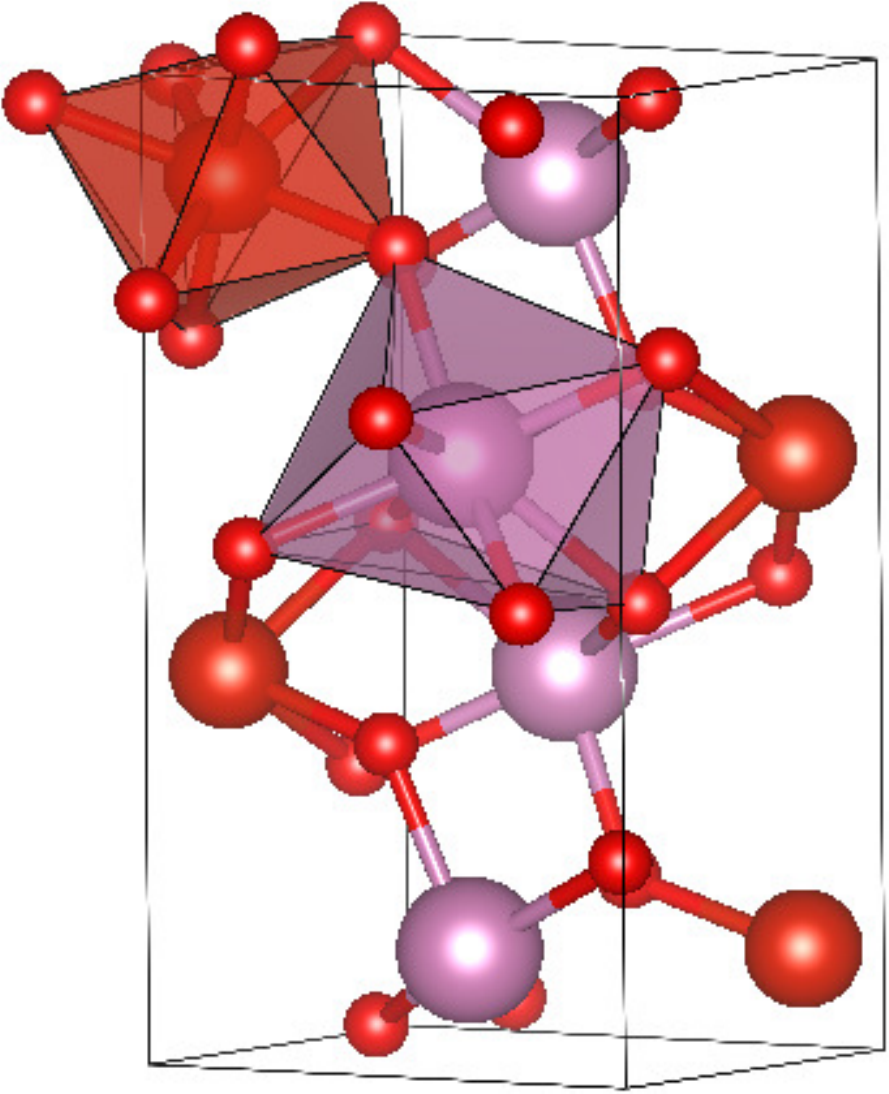} &
\includegraphics[width=3.1cm]{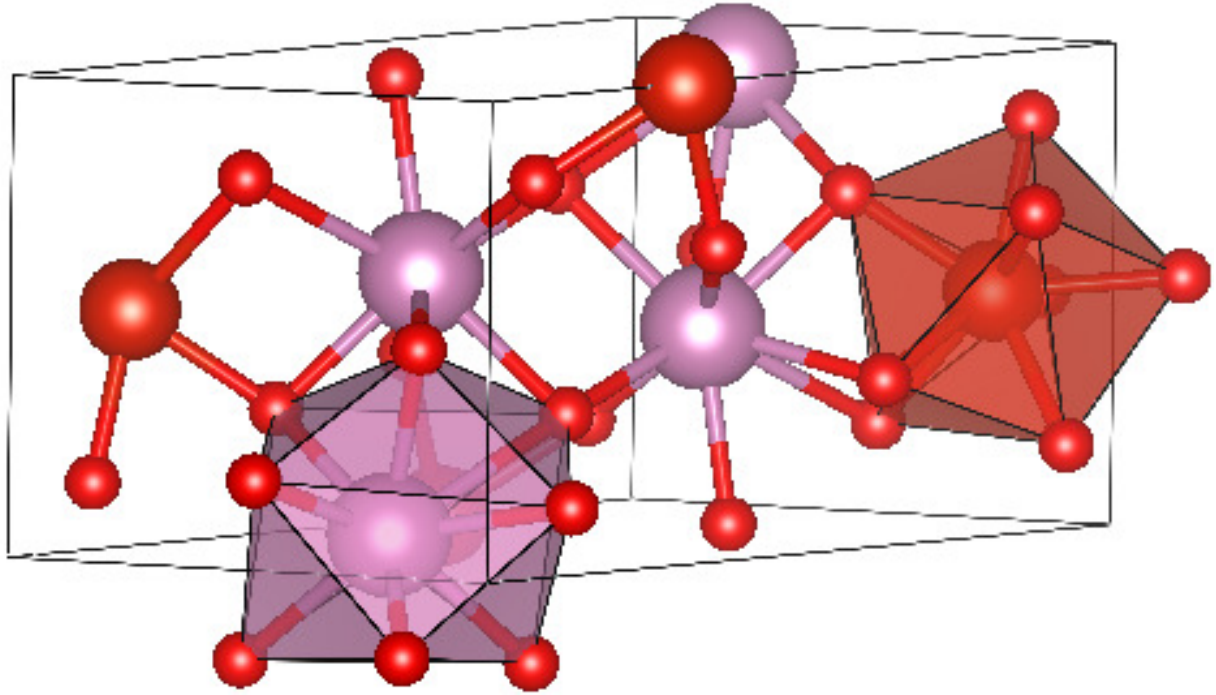}  \\
(a) & (b) & (c) & (d) & (e) & (f) \\
\end{tabular}
\caption{(Color online) Most relevant crystalline structures of InVO$_4$ 
polymorphs: (a) $\alpha$-MnMoO$_4$-type, (b) CrVO$_4$-type, (c) wolframite, 
(d) CuWO$_4$-type, (e) raspite, and (f) AgMnO$_4$-type.} 
\label{fig:1}
\end{figure}

InVO$_4$ can be synthesized in different polymorphic forms depending on the 
preparation and temperature conditions: InVO$_4$-I which has the 
$\alpha$-MnMoO$_4$-type structure (SG: $C2/m$, No. 12, $Z$ = 8, CS: 
monoclinic), InVO$_4$-II (undetermined structure), and the InVO$_4$-III also 
identified as the CrVO$_4$-type 
structure.~\cite{Touboul1980,Katari2013,Errandonea2013,Touboul1980b,
Roncaglia1986,Touboul1995,Mondal2016} Besides, two high pressure (HP) phases 
have been reported: InVO$_4$-IV (undetermined structure) and 
InVO$_4$-V~\cite{Errandonea2013} that
has the characteristic wolframite structure of  several $A$WO$_4$ compounds  
(SG: $P2/c$, No. 13, $Z$ = 2, CS: monoclinic).~\cite{Ruiz2011} The high pressure 
phenomena and the studies of phase transitions driven by pressure in $AB$O$_4$ 
compounds are broadly reviewed in Refs.~\citenum{Errandonea2008}, 
\citenum{Grochala2007} and \citenum{Muj03}, to name a few. Figure~\ref{fig:1} 
(a)-(c) shows the crystal structure of these phases, with the coordination polyhedra 
of In and V depicted in each figure. The most representative works on this 
compound have been dedicated to describe the crystalline structure of the phases 
III and V, whereas there is a latent lack of information regarding phases I, II, and 
IV, due to their relative stability against to the others. 

Regarding phase V, the wolframite structure matches with the high pressure phase 
reported for other compounds with CrVO$_4$-type structure such as InPO$_4$ and 
TiPO$_4$.~\cite{Lopez2012b} Besides, this phase is considered as a natural 
high-pressure post-CrVO$_4$-type structure in the phase diagram proposed for these 
kind of compounds.~\cite{Baran1998} Although the study performed by Errandonea 
\textit{et al.}~\cite{Errandonea2013} is a novelty and present important information 
about the stability of this compound under pressure, other transitions could be 
observed at higher pressures than those reached in their experiments. Moreover, it has 
been demonstrated that theoretical studies can provide new insights about the stability 
of $ABX_4$ compounds under pressure,~\cite{Errandonea2008} which could help 
to explore and understand the behavior of vanadates beyond the wolframite phase.

As well as studies on the crystalline CrVO$_4$-type structure of InVO$_4$, there have 
been investigations dedicated to the study of other forms of this compound. In these 
efforts, it was reported the synthesis and characterization of InVO$_4$ nanotubes 
arrays.~\cite{Wang2007,Liu2008,Yi2010} It is interesting to mention that these arrays 
can present either the $\alpha$-MnMoO$_4$-type structure~\cite{Wang2007,Yi2010} 
or the orthorhombic CrVO$_4$-type structure,~\cite{Yi2010} whereas nanoribbons
synthesized by hydrothermal process present the orthorhombic one. On the other hand, 
it has been reported that nanofibers of InVO$_4$ can present both 
structures.~\cite{Song2012} These results show the importance of studying the known 
polymorphs of InVO$_4$.

In an effort to get a better understanding of the behavior of InVO$_4$ at equilibrium 
and under pressure, we carry out \textit{ab initio} calculations by considering first 
the well known experimental synthesized phases of this compound to determine the 
stability of each reported structure: InVO$_4$-I, -III, and -V phases were studied. A 
complete description of the stable phases is presented in conjunction with the 
experimental data from literature. After, we go further and we put forward other 
possible high pressure phases for a range of pressure beyond the one reached in the 
experiments performed by Errandonea \textit{et al.}~\cite{Errandonea2013} to 
determine the transition pressures at ambient temperature and the range of pressure 
stability of each phase by using the quasiharmonic approximation.  We also report the 
evolution of phonon frequencies for the most stable polymorphs as well as the phonon 
spectrum and phonon density of states (phonon DOS) of each phase, which in turns 
will help us to determine the stability of each phase. In an effort to complement our 
study, we performed the band structure calculations for the most representative phases
of InVO$_4$.

The paper is organized as follows: In the next section, we give a detailed description 
of the computational procedure. The description of the structure of the experimentally 
known polymorphs of InVO$_4$ is presented in Sec.~\ref{sIII-A}, while the phase 
transitions driven by pressure are on Sec.~\ref{sIII-B}. The study of vibrational and 
electronic properties of  InVO$_4$ are shown in Sec.~\ref{sIII-C} and \ref{sIII-D}, 
respectively. Finally, the summary and conclusions are given in Sec.~\ref{sIV}.

\section{Computational details}\label{sII} 

Calculations of the total energy were performed within the framework of the density
functional theory (DFT) and the projector-augmented wave (PAW)~\cite{Blo94,Kre99} 
method as implemented in the Vienna \textit{ab initio} simulation package 
(VASP).~\cite{Kre93,Kre94a,Kre96a,Kre96b} A plane-wave energy cutoff of 520 eV
was used to ensure a high precision in our calculations. The exchange-correlation 
energy was described within the generalized gradient approximation (GGA) in the 
AM05~\cite{Armiento2005,Mattsson2009,Mattsson2008} formulation. 

The Monkhorst-Pack scheme was employed to discretize the Brillouin-zone (BZ) 
integrations~\cite{Mon76} with meshes 3$\times$3$\times$3, 4$\times$3$\times$3, 
4$\times$4$\times$2, 4$\times$4$\times$4, and 4$\times$4$\times$2, which 
correspond to sets of 10, 8, 4, 16, and 8, special \textit{k}-points in the irreducible BZ 
for the $\alpha$-MnMoO$_4$-type structure, CrVO$_4$-type structure, scheelite, 
wolframite, raspite (SG: $P2_1/a$, No. 14, $Z$=4, CS: monoclinic; this phase is the 
low temperature monoclinic dimorph of stolzite), and AgMnO$_4$-type structure (SG:
$P2_1/n$, No. 14, $Z$ = 4, CS: monoclinic), respectively. For the other phases 
considered in the high-pressure regime, we use the most suitable mesh for each case. 
In the relaxed equilibrium configuration, the forces are less than 2 meV/\AA\ per atom 
in each of the Cartesian directions. This high degree of convergence is required for the
calculations of vibrational properties using the direct force constant approach.~\cite{Par} 
The phonon DOS,  has been obtained from the calculation of the phonons in the whole 
BZ wih a supercell 2$\times$2$\times$2 times the conventional unit cell by using the 
PHONON software.~\cite{Par} The calculations of phonon spectrum were done for 
several volume points: 8 (for a range of pressure from $\approx$0 to 7 GPa), 11 (from 
3.6 to 35 GPa), 9 (from 22 to 51 GPa), and 6 (from 39.7 to 62 GPa) for the 
CrVO$_4$-type structure, wolframite, raspite, and the AgMnO$_4$-type structure, 
respectively. Temperature effects and zero-point energy have been included within the
quasiharmonic approximation~\cite{Baroni2001} through the calculation of the 
vibrational free energy, the method is well explained on references 
\citenum{Cazorla2009} and \citenum{Cazorla2015}. The phase transitions at 300 K 
were obtained analyzing the Gibbs free energy for the phases under study. For the 
electronic structure the optimized crystal structures were used with a larger set of 
$k$-points. We also used the hybrid 
HSE06~\cite{Heyd2003,Heyd2004,Heyd2006,Pair2006} exchange-correlation 
functional to calculate the electronic band structure and the electronic density of 
states. For these calculations we performed a full reoptimization of the structures 
obtained with the AM05 exchange correlation functional. In general, we found a 
difference of less than 1 GPa between the calculations performed with AM05 and 
HSE06 functionals for a specific volume.

\section{Results and Discussion}\label{sIII} 
\subsection{Structure of experimentally known polymorphs of InVO$_4$}\label{sIII-A} 

According to the literature, InVO$_4$ has been successfully synthesized and 
characterized in three different polymorphs, the $\alpha$-MnMoO$_4$-type 
(InVO$_4$-I),~\cite{Roncaglia1986,Touboul1995} the CrVO$_4$-type structure 
(InVO$_4$-III),~\cite{Errandonea2013,Katari2013} and, most recently, the wolframite 
(InVO$_4$-V).~\cite{Errandonea2013} While phases I and III crystallize at ambient 
pressure, phase V was  recently identified just under pressure.~\cite{Errandonea2013} 
Also, lately \textit{ab initio} calculations were used to study phases III and 
V.~\cite{Mondal2016} In order to determine the crystal structure and the relative 
stability of these phases at ambient and high pressure, we carry out the calculations 
of the simulations of these phases at different volumes. 

\begin{figure}[htb]
\centering
\begin{tabular}{cc}
\includegraphics[height=9.cm]{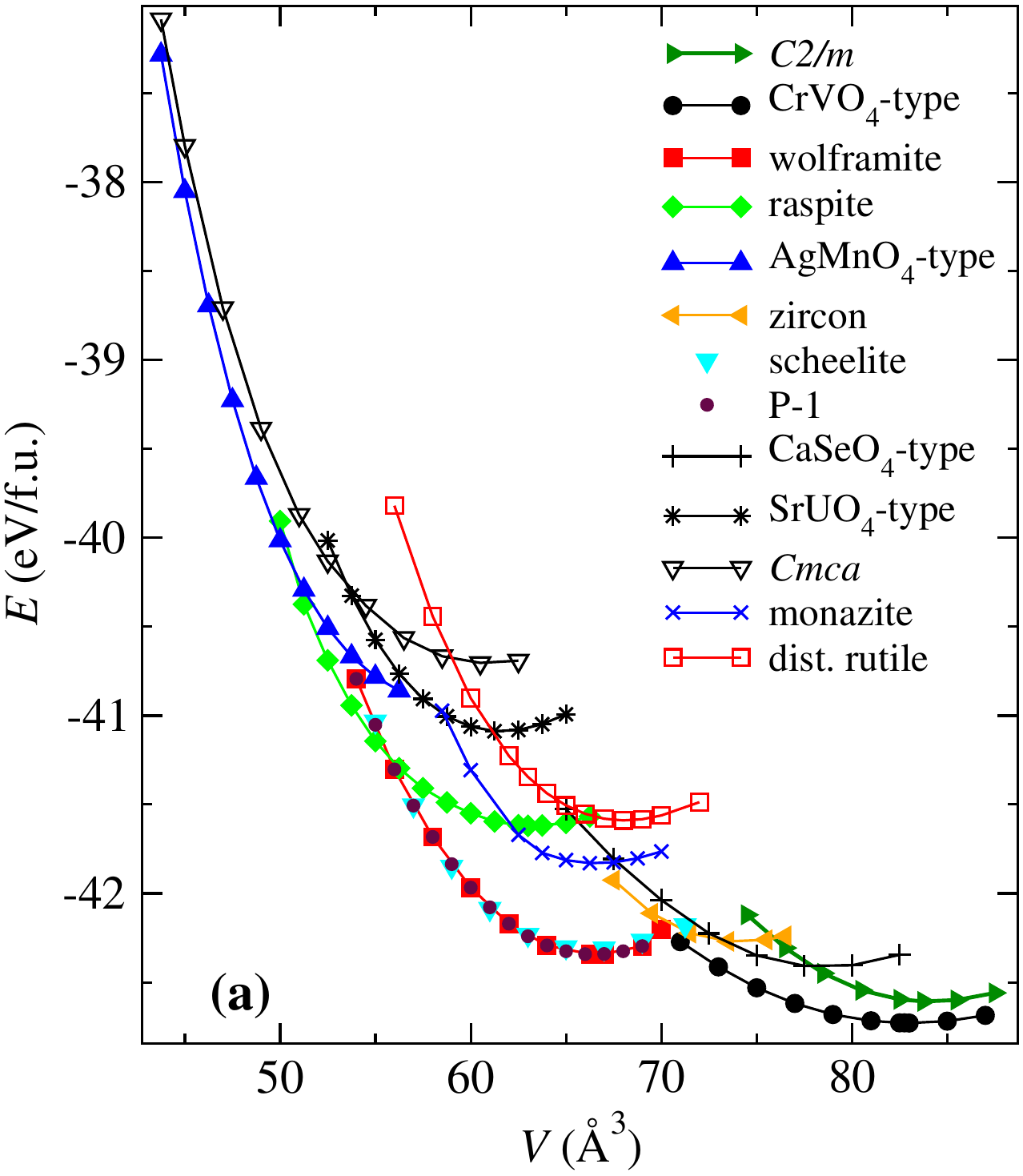}  &
\includegraphics[height=9.cm]{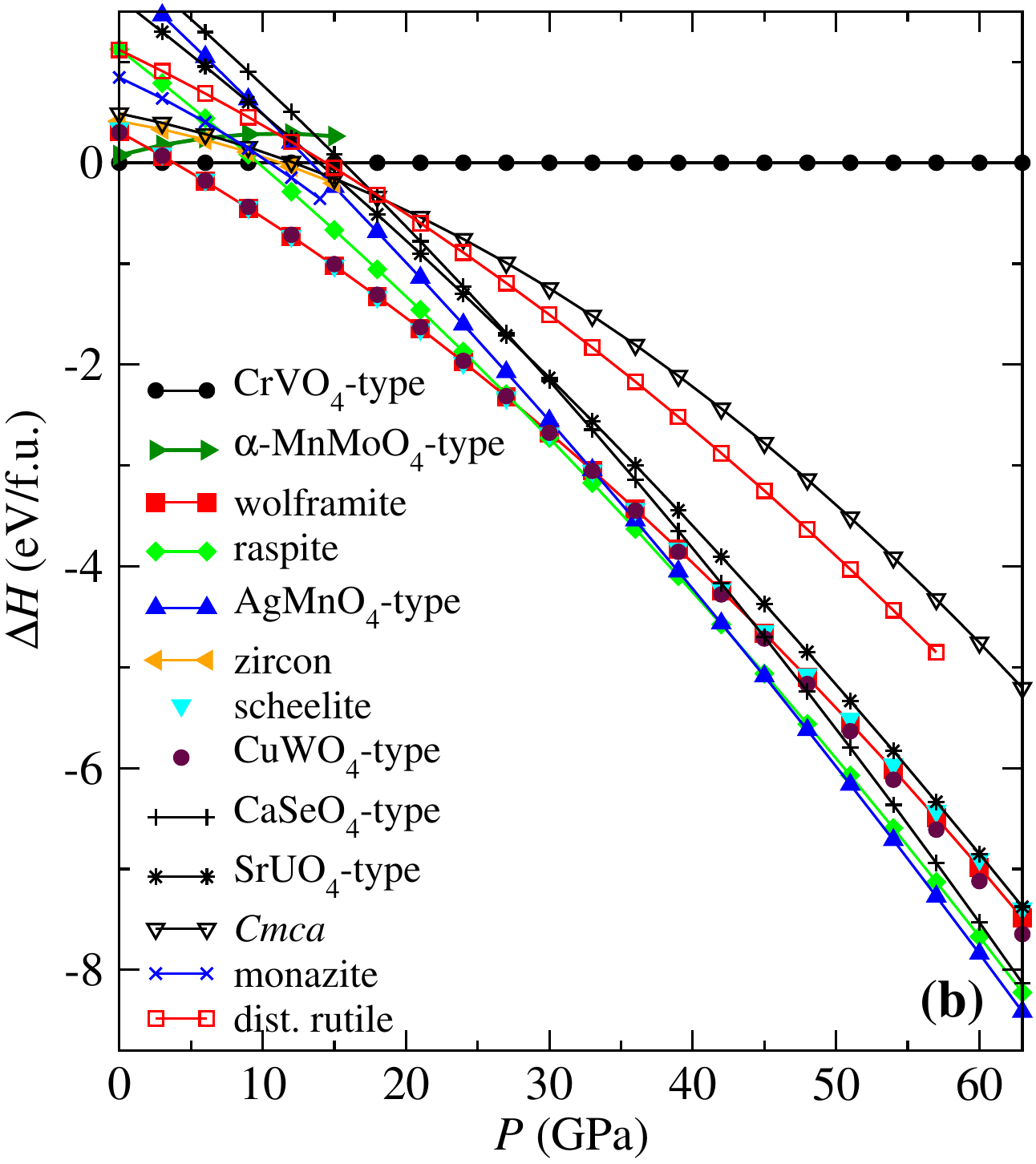}  \\
\end{tabular}
\caption{(Color online) (a) Calculated total energy-volume curves for the studied 
polymorphs of InVO$_4$ and (b) pressure dependence of enthalpy for the calculated 
phases of InVO$_4$.}
\label{fig:2}
\end{figure}

The equilibrium volume and unit-cell parameters were calculated by minimizing the 
crystal total energy for different volumes allowing to relax the internal atomic 
positions and lattice parameters. The volume-energy data were fitted with a third-order 
Birch-Murnaghan equation of state (EOS).~\cite{Birch47} As is shown in 
Fig.~\ref{fig:2} (a) and (b) the lowest energy structure of InVO$_4$ belongs to the 
CrVO$_4$-type structure followed by the $\alpha$-MnMoO$_4$-type structure 
and wolframite. Table~\ref{table:1} presents the optimized structural parameters 
obtained from our calculations, the theoretical results from Mondal 
\textit{et al.}~\cite{Mondal2016} and the experimental values published in the 
literature~\cite{Errandonea2013,Roncaglia1986,Touboul1995,Mondal2016} for 
comparison. For each phase, the pressure at which the values have been taken is 
indicated. As expected from a GGA exchange correlation functional, there is a small
overestimation of the calculated equilibrium volume with respect to experimental values. 
In the present case there is a good agreement with a difference of less than 1.5\% in the 
lattice parameters with respect to experimental values. Note that the corresponding 
difference between the results from Mondal \textit{et al.}~\cite{Mondal2016} and the
experimental data from Ref.~\citenum{Errandonea2013} is of the order of 3\%, which
represents a difference in the equilibrium volume of $\approx$7.4\%. Also, our results, 
of bulk modulus ($B_0$) and bulk pressure derivative ($B_0$'), are in better agreement 
than those from Mondal \textit{et al.}~\cite{Mondal2016} with respect to the 
experimental results from Ref.~\citenum{Errandonea2013}. For wolframite, the 
experimental parameters were obtained above 8 GPa, since the ambient pressure values
obtained from experiments are unknown due to the mixture of phases III and V 
observed once pressure is released from 23.9 GPa.~\cite{Errandonea2013} In the 
next section we will deal with the stability of these phases under pressure.

\setlength{\tabcolsep}{2pt}
\begin{table*}[t!]
\caption
{Structural parameters and bulk properties of experimentally known polymorphs of 
InVO$_4$ at ambient pressure and high pressures. Where $a$, $b$, and $c$ are the 
lattice parameters, $V$ is the volume ($V$/f.u. appear on parenthesis), $B$ the bulk 
modulus, and $B_0$' the pressure derivative of bulk modulus.}
\begin{tabular}{lccccccccccc}
\hline
& \multicolumn{3}{c}{$\alpha$-MnMoO$_4$-type} & \multicolumn{3}{c}{CrVO$_4$-type} & \multicolumn{3}{c}{wolframite} & \multicolumn{2}{c}{CuWO$_4$-type} \\
\cline{2-4} \cline{5-7} \cline{8-10} \cline{11-12}
& DFT\textsuperscript{\emph{a}}  & Exp.~\cite{Touboul1995} & Exp.~\cite{Roncaglia1986} 
& DFT\textsuperscript{\emph{a}}  & DFT~\cite{Mondal2016}   & Exp.~\cite{Errandonea2013} 
& DFT\textsuperscript{\emph{a}}  & DFT~\cite{Mondal2016}   & Exp.~\cite{Errandonea2013} 
& DFT\textsuperscript{\emph{a}}  & Exp.~\cite{Errandonea2013}  \\
\hline
$P$ (GPa)      &  atm.   &  atm.     & atm.       &  atm.  & atm.  & 0.8      &   8.5  & 6.0    &  8.2     & 8.57   & 8.2      \\
$a$ (\AA)      & 10.3516 & 10.271 & 10.49  & 5.7547 & 5.816 & 5.738 & 4.7009 & 4.776  & 4.714 & 4.6996 & 4.714 \\
$b$ (\AA)      &  9.4700 &  9.403 & 9.39    & 8.6168 & 8.739 & 8.492 & 5.5197 & 5.588  & 5.459 & 4.8840 & 5.459 \\
$c$ (\AA)      &  7.0863 &  7.038 & 7.12    & 6.6751 & 6.775 & 6.582 & 4.8849 & 4.958  & 4.903 & 5.5220 & 4.903 \\
$\alpha$ (deg.)&         &           &            &        &       &          &        &        &          & 90.021 & 90.2  \\
$\beta$ (deg.) & 104.81  & 105.08 & 105.1   &        &       &          & 92.62  & 92.89  & 93.8  & 90.030 & 93.8  \\
$\gamma$ (deg.)&         &           &            &        &       &          &        &        &          & 92.588 & 90.2  \\
$Z$            &  8      &   8       &  8         &  4     &  4    &   4      &   2    &   2    &   2      & 2      & 2        \\
$V$ (\AA$^3$)&  671.6  & 656.3  & 677.4      & 331.0  &344.37 &  320.72  & 126.62 & 132.17 & 125.89& 126.62 & 125.89   \\
                             & (83.95) & (82.04) & (84.67) & (82.75) & ( 86.09) & (80.18) & (63.31) & (66.08) & (62.94) & (63.31) & (62.94)       \\
$B_0$ (GPa)    &  120.9  &            &   & 71.0  & 76.47 &  69   &  166.1 & 183.0  & 168   & 166.1  & 168   \\
$B_0$'         &   4.52  &            &   & 4.0   &  3.0  &   4.0    &  4.26  &  6.0   &  4.0     & 4.26   & 4.0 \\
\hline
\end{tabular}
\textsuperscript{\emph{a}} This work.
\label{table:1}
\end{table*}

\setlength{\tabcolsep}{5pt}
\begin{table*}[ht!]
\caption
{Wyckoff positions of the experimentally known polymorphs of InVO$_4$ for values of 
Table~\ref{table:1}.}
\begin{tabular}{lcccccccccc}
\hline
\multicolumn{11}{c}{$\alpha$-MnMoO$_4$-type} \\ 
&      & \multicolumn{3}{c}{DFT} & \multicolumn{3}{c}{Exp.~\cite{Touboul1995}} \\
\cline{3-5} \cline{6-8} 
 Atom  & WP   & $x$    & $y$    & $z$    & $x$       & $y$       & $z$       \\
\hline
           In$_1$& 4$h$ & 0      & 0.1870 & 0.5    & 0          & 0.1875 & 0.5        \\
           In$_2$& 4$i$ & 0.7892 & 0      & 0.1317 & 0.7907  & 0         & 0.1313  \\
           V$_1$ & 4$g$ & 0      & 0.2520 & 0      & 0          & 0.2563 & 0          \\
           V$_2$ & 4$i$ & 0.2737 & 0      & 0.4061 & 0.2737  & 0         & 0.402   \\
           O$_1$ & 4$i$ & 0.1410 & 0      & 0.5394 & 0.141   & 0.5       & 0.541   \\
           O$_2$ & 4$i$ & 0.7915 & 0      & 0.8374 & 0.789   & 0         & 0.831   \\
           O$_3$ & 8$j$ & 0.6324 & 0.1496 & 0.1074 & 0.636   & 0.153  & 0.107   \\        
           O$_4$ & 8$j$ & 0.0481 & 0.1553 & 0.8127 & 0.044   & 0.161  & 0.808  \\
           O$_5$ & 8$j$ & 0.1393 & 0.3512 & 0.5252 & 0.136   & 0.350  & 0.522   \\
\\
\multicolumn{11}{c}{CrVO$_4$-type}  \\ 
&      & \multicolumn{3}{c}{DFT} & \multicolumn{3}{c}{DFT~\cite{Mondal2016}} & \multicolumn{3}{c}{Exp.~\cite{Errandonea2013}} \\
\cline{3-5} \cline{6-8} \cline{9-11}
Atom  & WP   & $x$    & $y$    & $z$    & $x$   & $y$   & $z$   & $x$       & $y$       & $z$       \\
\hline
       In    & 4$a$ & 0      & 0      & 0      & 0     & 0     & 0     & 0         & 0         & 0         \\
       V     & 4$c$ & 0      & 0.3581 & 0.25   & 0     & 0.358 & 0.25  & 0         & 0.3617 & 0.25      \\
       O$_1$ & 8$g$ & 0.2575 & 0.4730 & 0.25   & 0.256 & 0.473 & 0.25  & 0.2568 & 0.4831 & 0.25      \\
       O$_2$ & 8$f$ & 0      & 0.7543 & 0.9536 & 0     & 0.753 & 0.953 & 0         & 0.7492 & 0.9573 \\
\\
\multicolumn{11}{c}{wolframite} \\
&      & \multicolumn{3}{c}{DFT} & \multicolumn{3}{c}{DFT~\cite{Mondal2016}} & \multicolumn{3}{c}{Exp.~\cite{Errandonea2013}}  \\
\cline{3-5} \cline{6-8} \cline{9-11}
Atom  & WP   & $x$    & $y$    & $z$    & $x$   & $y$   & $z$   & $x$      & $y$      & $z$      \\
\hline
               In    & 2$f$ & 0.5    & 0.7040 & 0.25   & 0.5   & 0.704 & 0.25  & 0.5      & 0.711 & 0.25  \\
               V     & 2$e$ & 0      & 0.1814 & 0.25   & 0     & 0.181 & 0.25  & 0        & 0.159 & 0.25      \\
               O$_1$ & 4$g$ & 0.2128 & 0.9086 & 0.4574 & 0.21  & 0.91  & 0.46  & 0.214 & 0.861 & 0.492  \\
               O$_2$ & 4$g$ & 0.2499 & 0.3834 & 0.3885 & 0.248 & 0.382 & 0.386 &0.241 & 0.407 & 0.399  \\
\\
\multicolumn{11}{c}{CuWO$_4$-type} \\
&      & \multicolumn{3}{c}{DFT} & \multicolumn{3}{c}{Exp.~\cite{Errandonea2013}}  \\
\cline{3-5} \cline{6-8} 
Atom  & WP   & $x$    & $y$    & $z$    & $x$   & $y$   & $z$       \\
\hline
               In    & 2$i$ & 0.4999 & 0.7498 & 0.2964 &   0.5   & 0.711 & 0.25  \\
               V     & 2$i$ & 0.0004 & 0.7501 & 0.8188 &   0     & 0.159 & 0.25  \\
               O$_1$ & 2$i$ & 0.7871 & 0.5429 & 0.0915 &  0.214 & 0.861 & 0.491 \\
               O$_2$ & 2$i$ & 0.2127 & 0.9569 & 0.0916 &  0.786 & 0.861 & 0.492 \\
               O$_3$ & 2$i$ & 0.7501 & 0.6117 & 0.6168 &  0.242 & 0.407 & 0.399 \\
               O$_4$ & 2$i$ & 0.2500 & 0.8884 & 0.6168 &  0.758 & 0.407 & 0.101 \\
\hline
\end{tabular}
\label{table:2}
\end{table*}

\setlength{\tabcolsep}{5pt}
\begin{table}[ht!]
\caption
{Interatomic bond distances, In-O and V-O for the experimental known polymorphs of InVO$_4$. 
The experimental data were taken from Refs.~\citenum{Touboul1995} and \citenum{Errandonea2013}.}
\begin{tabular}{lcclcc}
\hline
& \multicolumn{2}{c}{In-O (\AA)} & & \multicolumn{2}{c}{V-O (\AA)}   \\
\cline{2-3} \cline{5-6} 
& DFT & Exp. & & DFT & Exp. \\
\hline
\multicolumn{6}{c}{$\alpha$-MnMoO$_4$-type} \\
In$_1$-O$_1$($\times$2) & 2.2667 & 2.25 & V$_1$-O$_3$($\times$2) & 1.6714 & 1.64 \\
In$_1$-O$_4$($\times$2) & 2.1645 & 2.11 & V$_1$-O$_4$($\times$2) & 1.7848 & 1.78 \\
In$_1$-O$_5$($\times$2) & 2.0969 & 2.05 & V$_2$-O$_1$($\times$1) & 1.8544 & 1.87 \\
In$_2$-O$_1$($\times$1) & 2.2571 & 2.23 & V$_2$-O$_2$($\times$1) & 1.6856 & 1.59 \\
In$_2$-O$_2$($\times$1) & 2.0917 & 2.11 & V$_2$-O$_5$($\times$2) & 1.6763 & 1.69 \\
In$_2$-O$_3$($\times$2) & 2.1276 & 2.12 & $\langle$In-O$\rangle$ & 1.7256 & 1.71    \\
In$_2$-O$_4$($\times$2) & 2.1947 & 2.23 &   \\
$\langle$In-O$\rangle$  & 2.1708 & 2.155   &   \\
\\
\multicolumn{6}{c}{CrVO$_4$-type}   \\
In-O$_2$ ($\times$2)   & 2.1392 & 2.1483 & V-O$_2$ ($\times$2)  & 1.6690 & 1.6579  \\
In-O$_1$ ($\times$4)   & 2.1877 & 2.1623 & V-O$_1$ ($\times$2)  & 1.7824 & 1.7983  \\
$\langle$In-O$\rangle$ & 2.1755 & 2.1588    & $\langle$V-O$\rangle$& 1.7257 & 1.7281     \\
\multicolumn{6}{c}{CrVO$_4$-type~\cite{Mondal2016}}   \\
In-O$_2$ ($\times$2)   & 2.22 &  & V-O$_2$ ($\times$2)  & 1.68 &   \\
In-O$_1$ ($\times$4)   & 2.18 &  & V-O$_1$ ($\times$2)  & 1.79 &   \\
$\langle$In-O$\rangle$ & 2.193 &     & $\langle$V-O$\rangle$& 1.735 &      \\
\\
\multicolumn{6}{c}{wolframite}   \\
In-O$_1$ ($\times$2)   & 2.0610 & 2.0268 & V-O$_1$ ($\times$2)  & 1.7346 & 1.6730  \\
In-O$_2$ ($\times$2)   & 2.1316 & 2.1397 & V-O$_1$ ($\times$2)  & 2.0501 & 2.2166  \\
In-O$_2$ ($\times$2)   & 2.2459 & 2.2101 & V-O$_2$ ($\times$2)  & 1.8498 & 1.8861  \\
$\langle$In-O$\rangle$ & 2.1462 & 2.1255 & $\langle$V-O$\rangle$& 1.8782 & 1.9252     \\
\multicolumn{6}{c}{wolframite~\cite{Mondal2016}}   \\
In-O$_1$ ($\times$2)   & 2.11 &  & V-O$_1$ ($\times$2)  & 1.75 &   \\
In-O$_2$ ($\times$2)   & 2.18 &  & V-O$_1$ ($\times$2)  & 2.08 &   \\
In-O$_2$ ($\times$2)   & 2.29 &  & V-O$_2$ ($\times$2)  & 1.87 &   \\
$\langle$In-O$\rangle$ & 2.193 &  & $\langle$V-O$\rangle$& 1.90 &      \\
\\
\multicolumn{6}{c}{CuWO$_4$-type }   \\
In-O$_1$ ($\times$1)   & 2.0605 &           & V-O$_1$ ($\times$1)  & 1.7340      \\
In-O$_2$ ($\times$1)   & 2.0607 &           & V-O$_2$ ($\times$1)  & 1.7351      \\
In-O$_3$ ($\times$1)   & 2.1313 &           & V-O$_3$ ($\times$1)  & 1.8492      \\
In-O$_4$ ($\times$1)   & 2.1319 &           & V-O$_4$ ($\times$1)  & 1.8506      \\
In-O$_5$ ($\times$1)   & 2.2445 &           & V-O$_5$ ($\times$1)  & 2.0480      \\
In-O$_6$ ($\times$1)   & 2.2453 &           & V-O$_6$ ($\times$1)  & 2.0511      \\
$\langle$In-O$\rangle$ & 2.1457 &           & $\langle$V-O$\rangle$& 1.8780   \\
\hline
\end{tabular}
\label{table:3}
\end{table}

Reports related to the InVO$_4$-I phase are  limited to the work performed by Touboul 
\textit{et al.}~\cite{Touboul1980,Touboul1980b} and Roncaglia \textit{et 
al.}~\cite{Roncaglia1986} This phase has not attracted much attention, due to its  low 
stability. However, it is well known that this phase resembles the structure of 
CrVO$_4$-I~\cite{Touboul1995b} and $\alpha$-MnMoO4.~\cite{Abrahams1965} 
Therefore, a good description of this phase is needed, furthermore it could help to 
understand the $\alpha$-MnMoO4 compound and other molibdates with the same 
structure.~\cite{Sleight1968} The InVO$_4$-I phase has eight f.u. in the unit cell. 
In this structure there are two nonequivalent positions for In and V and five oxygen 
positions (Table~\ref{table:2}), which leads to several different In-O and V-O bond 
distances, as can be appreciated in Table~\ref{table:3}. The values for In-O (V-O) 
range from 2.092 (1.671) to 2.267 (1.854) \AA, in good agreement with experimental 
results.~\cite{Touboul1995} In this structure, In and V atoms are surrounded by six 
and four oxygen atoms, respectively, to form the irregular InO$_6$ octahedra and 
VO$_4$ tetrahedra (Figure~\ref{fig:1} a). The four InO$_6$ units share edges along 
$y$ and $z$ directions to form In$_4$O$_{16}$ clusters which share corners with 
sixteen VO$_4$ tetrahedras in such a way that layers of In$_4$O$_{16}$ clusters 
are formed separated by VO$_4$ units in the (20$\bar 1$) plane. The VO$_4$ units 
do not share corners with each other, but are connected to the corners of each InO$_6$ 
octahedra. We have to mention that the formation of these clusters was not observed 
in the other phases studied in this work.

The CrVO$_4$-type structure is the most studied phase of InVO$_4$. We just mention 
that the structure consist of edge-sharing InO$_6$ octahedra  along the $c$ direction, 
the chains are linked together with VO$_4$ tetrahedra (Figure 1-a). The tetrahedra and 
octahedra are more regular than in the $\alpha$-MnMoO4 structure as can be inferred 
from the In-O an V-O distances displayed in Table~\ref{table:3}. The apical distances 
of the InO$_6$ polyhedra are 2.1392 ($\times$2) \AA\ and the equatorial are 2.1877 
($\times$4) \AA. In this structure the VO$_4$ tetrahedra are not linked to one another. 
The In and V atoms occupy 4$a$ and 4$c$ positions, respectively, while there are two 
nonequivalent oxygen atoms in 8$g$ and 8$f$ positions, see Table~\ref{table:2}. As can 
be seen from Tables~\ref{table:1} to \ref{table:3} our results are in good agreement with
the experimental data reported in Refs.~\citenum{Katari2013} and 
\citenum{Errandonea2013}.

The structure of wolframite was widely studied for other compounds that crystallizes in 
this structure at ambient pressure, such as $AB$O$_4$ [$A$ = Mg, Mn, Fe, Co, Ni, $B$ 
= Mo, W].~\cite{Ruiz2010,Ruiz2011,Ruiz2012,Lopez2009,Sleight1972,Errandonea2008} 
Since the experimental data are reported above 8 GPa, our theoretical description of this 
phase will be done for the structure at 8.5 GPa. At this pressure there are remarkable 
differences between this phase and the others previously described. From bond distances 
in Table~\ref{table:3}, it can be deduced that polyhedra InO$_6$ and VO$_6$, 
Fig.~\ref{fig:1} (c), are more irregular than in phase III. The structure consist of 
alternating InO$_6$ and VO$_6$ octahedral units that share edges, forming zigzag 
chains building a close-packed structure.~\cite{Ruiz2011} These alternating chains are 
the reason of the highest bulk modulus of wolframite in comparison with phases I and III. 
In this structure, the VO$_6$ octahedra are less compressible than the InO$_6$ 
polyhedra.~\cite{Errandonea2008} This topic will treated in more detail in the next section.

In order to correlate our results with the experimental ones,~\cite{Errandonea2013} we 
used the coordinates of the reported triclinic $P\bar1$ structure,~\cite{Errandonea2013}
which could be described as a CuWO$_4$-type structure (SG: $P\bar 1$, No. 2, $Z$ = 
2, CS: triclinic),~\cite{Ruiz2010b} to calculate the stability of this phase against the 
wolframite. As seen in Fig.~\ref{fig:2} (a) wolframite and CuWO$_4$-type structure are
energetically competitive, as a mater of fact it seems that CuWO$_4$-type structure is a
distortion of the wolframite one. According to Table~\ref{table:1} to \ref{table:3} this
distortion is small, so small that the difference in the results for phase transitions and 
phonons are almost negligible, as we will see in the next sections.

\subsection{Phase transitions}\label{sIII-B}

To the best of our knowledge there are very few studies about the stability of InVO$_4$ 
under extreme conditions of temperature~\cite{Katari2013} and 
pressure.~\cite{Errandonea2013,Mondal2016} The X-ray diffraction pattern of 
InVO$_4$ in the CrVO$_4$-type structure  was reported  from ambient temperature to 
1023 K.~\cite{Katari2013} No phase transition in this range of temperature was reported, 
only a smooth increase in the unit-cell parameters and a significantly higher conductivity
above 723 K were observed. On the other hand, according to the high pressure studies 
from Ref.~\citenum{Errandonea2013} the InVO$_4$ undergoes a phase transition at 7 
GPa from the CrVO$_4$-type structure to the novel polymorph wolframite, which has 
been designed as phase V of this compound. This phase transition was accompanied by a
volume reduction of 14\%. Another phase between III and V was also observed in the 
experiments performed by Errandonea \textit{et al.}~\cite{Errandonea2013} This phase 
IV could not be well described because it appears as a minority phase in the X-ray patterns 
in a small range of pressure coexisting with phases III and V. As is explained in 
Ref.~\citenum{Errandonea2013}, phase IV was never observed as a pure phase. 

On the other hand, first-principles calculations were used to study the phase transition 
from CrVO$_4$-type structure to wolframite.~\cite{Mondal2016} Where a phase 
transition pressure of 5.6 GPa was reported. In their work Mondal \textit{et al.} 
reported the pressure evolution of lattice parameters, volume and interatomic bond 
distances for both phases up to a pressure of 14 GPa.~\cite{Mondal2016}

Regarding to the high pressure behavior of other $AB$O$_4$ compounds with 
CrVO$_4$-type structure, it has been reported that TiPO$_4$ and InPO$_4$ follow 
the phase transition sequence CrVO$_4$-type $\rightarrow$ zircon $\rightarrow$ 
scheelite $\rightarrow$ wolframite.~\cite{Lopez2012b} The first two transitions 
were also observed in TiSiO$_4$.~\cite{Gracia2009} Besides, CaSO$_4$ undergoes 
the next phase transitions: CrVO$_4$-type $\rightarrow$ monazite $\rightarrow$ 
barite (SG: $Pbnm$, No. 62, $Z$ = 4, CS: orthorhombic) $\rightarrow$ 
scheelite.~\cite{Gracia2012} While CaSeO$_4$, which crystallizes in an structure 
with space group very close to the CrVO$_4$-type structure (SG: $Cmca$, No. 64, 
$Z$ = 4, CS: orthorhombic), experiments the phase transition sequence $Cmca$ 
$\rightarrow$ scheelite $\rightarrow$ AgMnO$_4$-type structure.~\cite{Lopez2015}.
Otherwise, much less studies have been devoted to study FeVO$_4$ and CrVO$_4$ 
under pressure.~\cite{Young1962,Laves1964} 

Going beyond the pressures range achieved in the experiments conducted by 
Errandonea \textit{et al.},~\cite{Errandonea2013} we analyze the high pressure 
behavior of InVO$_4$ and we consider several possible structures in addition to the 
known polymorphs reported in the literature. To make the selection of candidates 
phases we take into consideration the north-east trend followed by other compounds 
in the Bastide's diagram~\cite{Errandonea2008} and previous studies performed in 
$AB$O$_4$ compounds.~\cite{Gracia2009,Gracia2012,Lopez2012b,Lopez2015,
Young1962,Laves1964,Baran1998} Among the selected structures are the zircon, 
which is the structure of other vanadates such as HoVO$_4$~\cite{Garg2014} and 
CeVO$_4$,~\cite{Panchal2011} and the scheelite,~\cite{Lopez2011} which has 
been reported as a high pressure phase of several vanadates~\cite{Panchal2011b} 
and some compounds that crystallizes in the CrVO$_4$-structure~\cite{Lopez2012b,
Gracia2009,Gracia2012} and in the $Cmca$-type structure.~\cite{Lopez2015} 

We have also considered the monazite phase, this structure occurs frequently 
among $AB$O$_4$ compounds.~\cite{Clavier2011} In particular, this structure 
is a prototype of high pressure phase of CrVO$_4$-type 
compounds~\cite{Baran1998} and has been reported as a high-pressure phase of 
PrVO$_4$,~\cite{Errandonea2013b} CaSO$_4$,~\cite{Gracia2012} and 
CaSeO$_4$.~\cite{Lopez2015} Other phases studied such as raspite and 
AgMnO$_4$-type were found as a post-scheelite phase of 
CaSO$_4$,~\cite{Crichton2005} CaSeO$_4$,~\cite{Lopez2015} and 
SrCrO$_4$.~\cite{Gleissner2016} Some other post-scheelite~\cite{Lopez2012b} 
phase were considered as the SrUO$_4$-type (SG: $Pbcm$, No. 57, $Z$= 4, CS:
orthorhombic) and the distorted (dist.) rutile (SG: $Cmmm$, No. 65, $Z$ = 2, CS: 
orthorhombic), is a prototype of high pressure phase for CrVO$_4$-type 
compounds.~\cite{Baran1998} Two different structures with $Cmca$ space 
group (No. 64) were also included in our study, one corresponds to the 
polymorphous reported of CaSeO$_4$ with $Z$ = 4,~\cite{Lopez2015} and the 
other one with 8 f.u. in the unit cell corresponds to the high-pressure phase reported 
for some $A$WO$_4$ compounds.~\cite{Errandonea2008} To compare our results 
with the experimental data, we take into account the CuWO$_4$-type 
structure.~\cite{Ruiz2010b} For completeness, we also include the study of the 
possible decomposition of InVO$_4$ under pressure to form InO + VO$_3$, in 
order to found if this matches with the experimental data of phase IV. 

\begin{figure}[h]
\centering
\begin{tabular}{c}
\includegraphics[width=8.5cm]{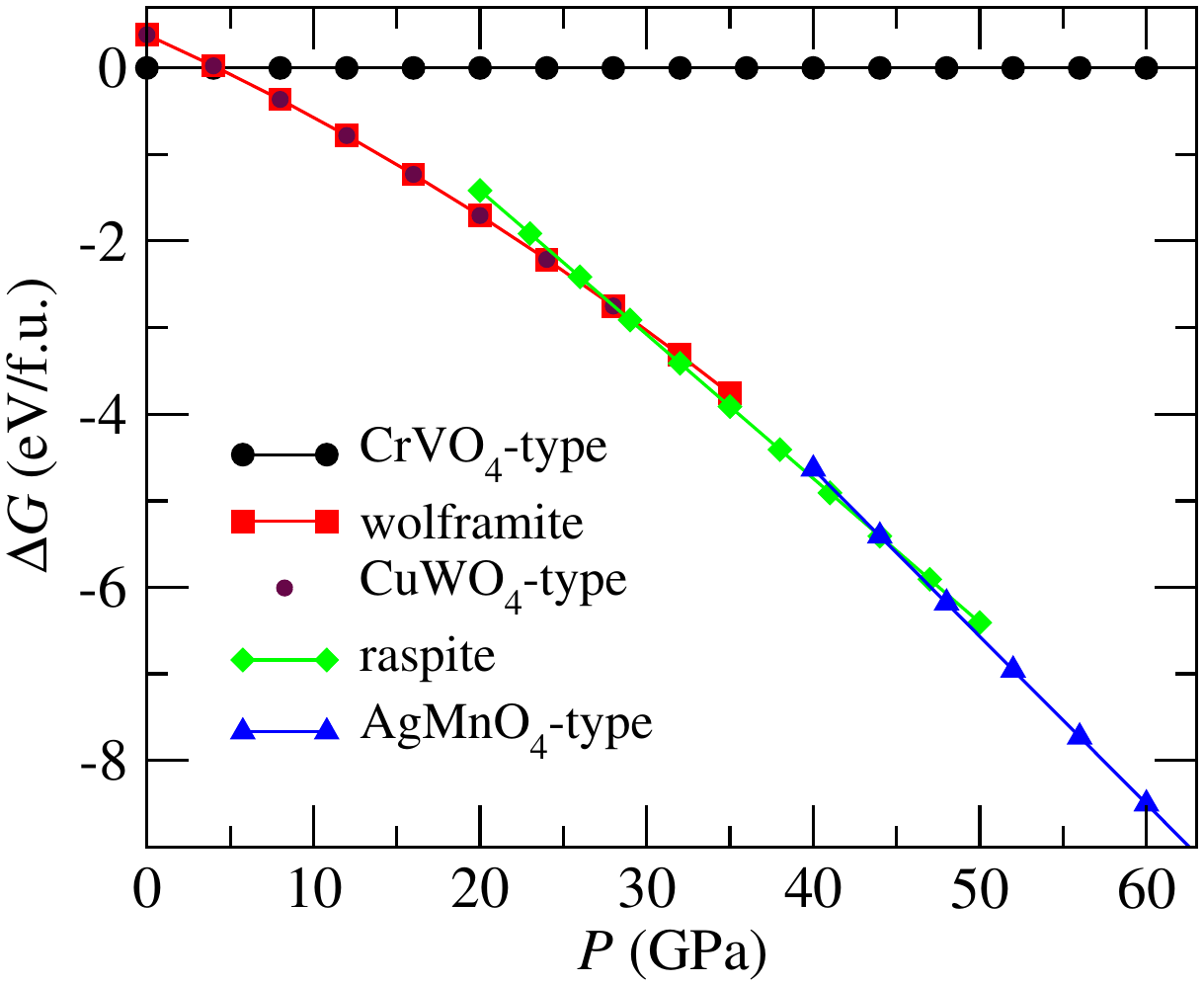}  
\end{tabular}
\caption{(Color online) Pressure dependence of Gibbs free energy (at 300 K)
for the most stable phases of InVO$_4$.}
\label{fig:3}
\end{figure}

The calculated energy-volume curves for the mentioned polymorphs of InVO$_4$ 
are illustrated in Fig.~\ref{fig:2} (a). The relative stability and coexistence pressures 
of these phases can be extracted by the common-tangent construction.~\cite{Muj03} 
The pressure-enthalpy diagram for the considered structures shows that, besides the 
known polymorphs of InVO$_4$, only the raspite, AgMnO$_4$-type, and the 
CuWO$_4$-type are competitive in the high-pressure range, see Fig.~\ref{fig:2} (b). 
Hence, we only calculated the Gibbs energy for these structures. We have to 
remember that the calculation of Gibbs free energy by the procedure described in 
Section \ref{sII} is computationally expensive, reason for which we only get the 
Gibbs free energy for the most representative phases of InVO$_4$. Figure~\ref{fig:3} 
shows the pressure evolution of the Gibbs energy difference, $\Delta G$, for the most 
representative phases at 300 K.

\begin{figure}[t!]
\centering
\begin{tabular}{c}
\includegraphics[width=8.5cm]{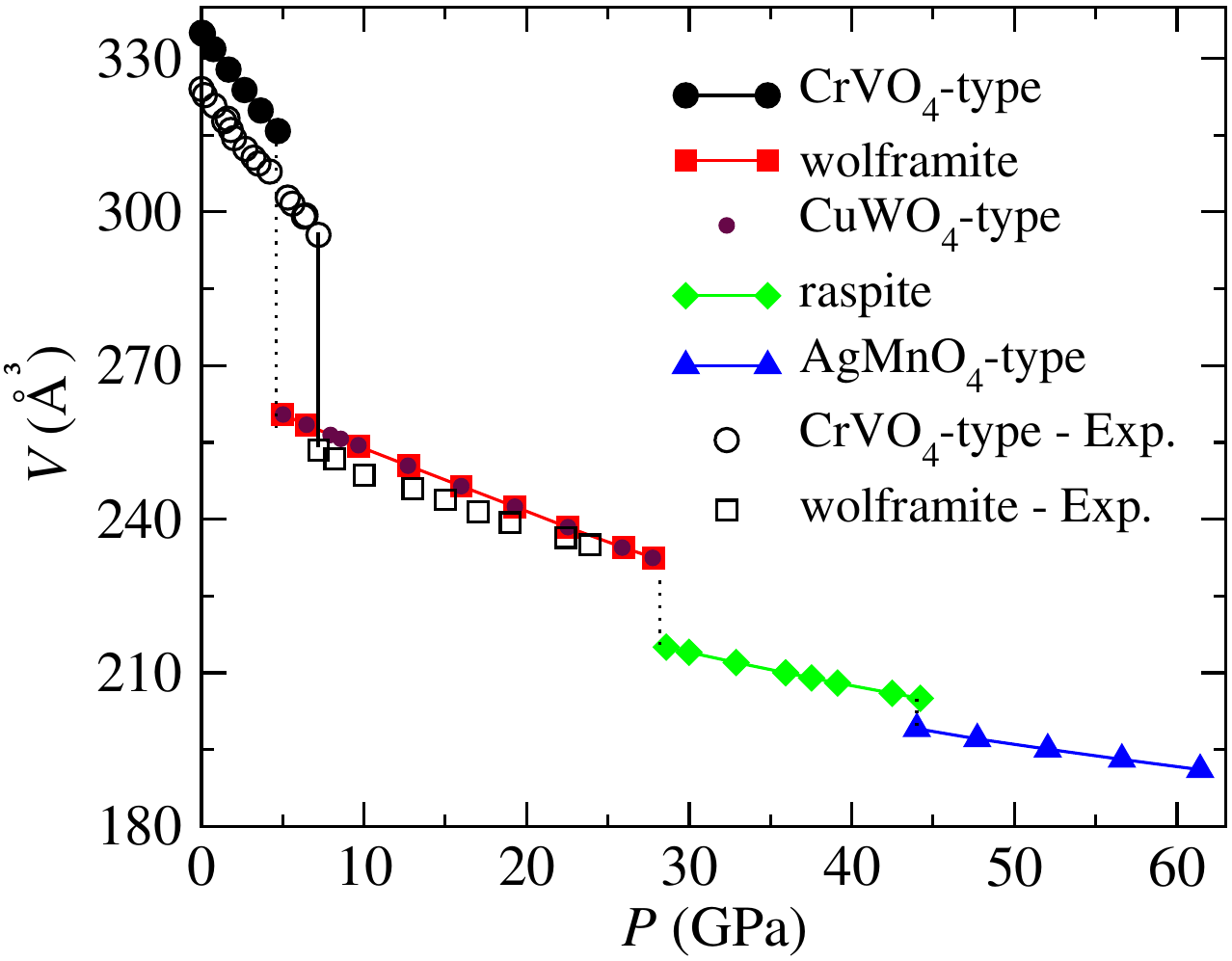}  
\end{tabular}
\caption{(Color online) Volume curves as function of pressure for the most stable 
polymorphs of InVO$_4$. The experimental data was taken from 
Ref.~\citenum{Errandonea2013}. Vertical doted (continuous) lines show the path 
followed in the phase transitions obtained from DFT calculations (experiments).}
\label{fig:4}
\end{figure}

\begin{figure}[t!]
\centering
\begin{tabular}{r}
\includegraphics[width=8.41cm]{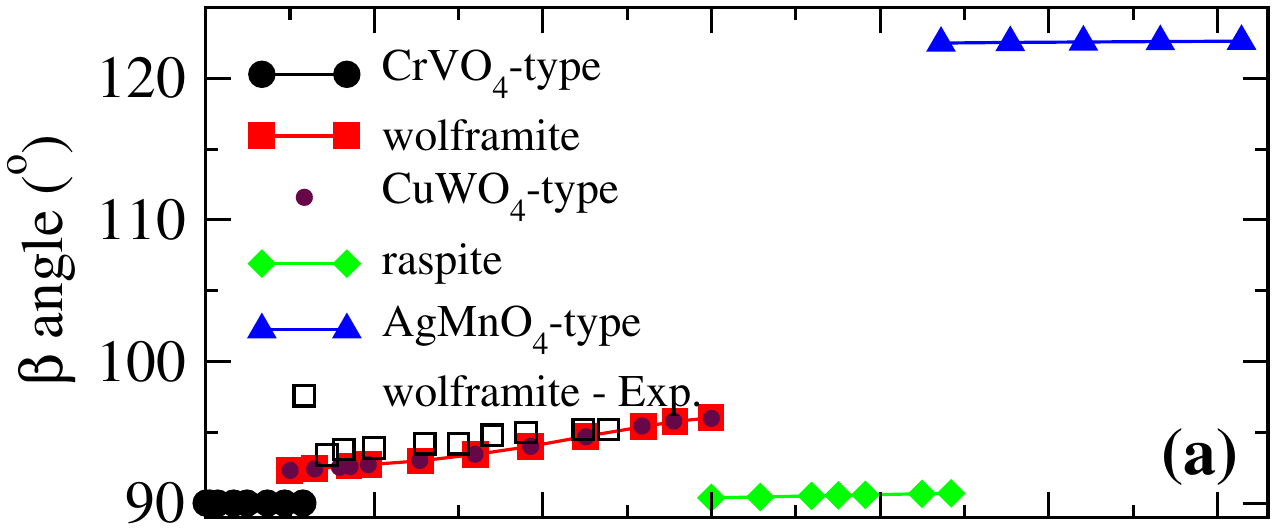}  \\    
\includegraphics[width=8.00cm]{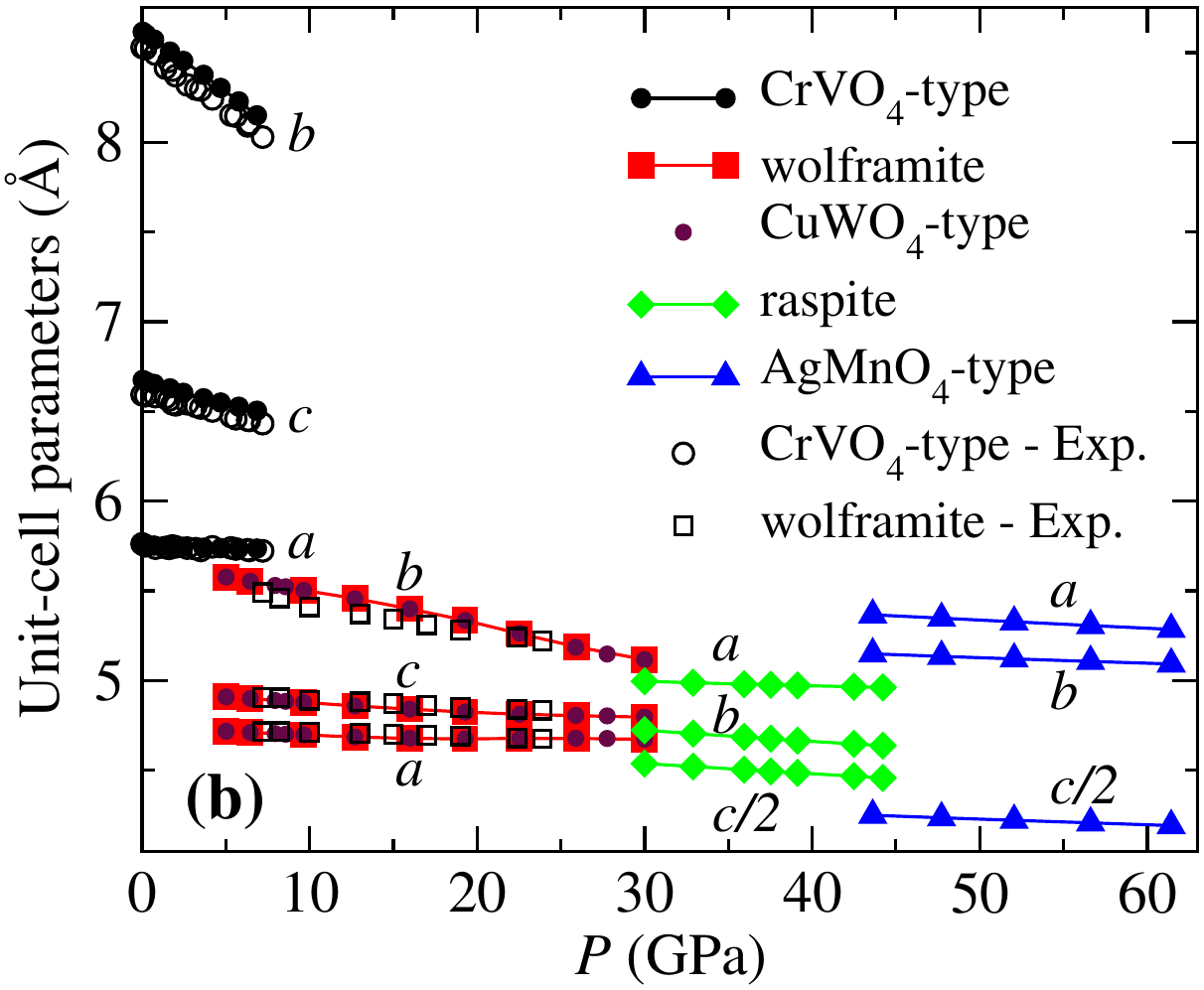}    
\end{tabular}
\caption{(Color online) Pressure dependence of (a) $\beta$ angle and (b) lattice 
parameters for the studied polymorphs of InVO$_4$. The experimental data was 
taken from Ref.~\citenum{Errandonea2013}.}
\label{fig:5}
\end{figure}

\begin{figure}[h!]
\centering
\begin{tabular}{r}
\includegraphics[width=8.5cm]{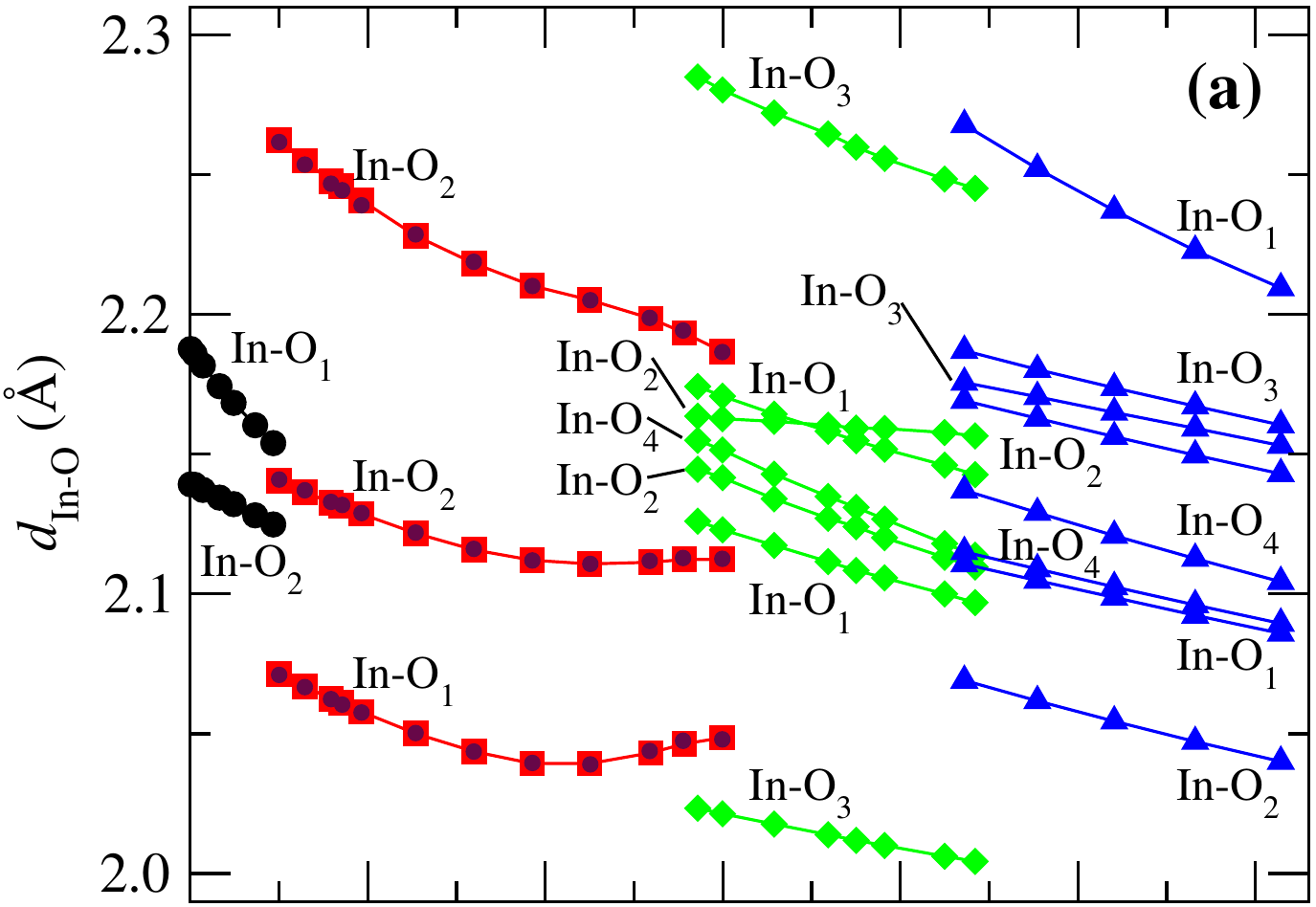}  \\    
\includegraphics[width=8.5cm]{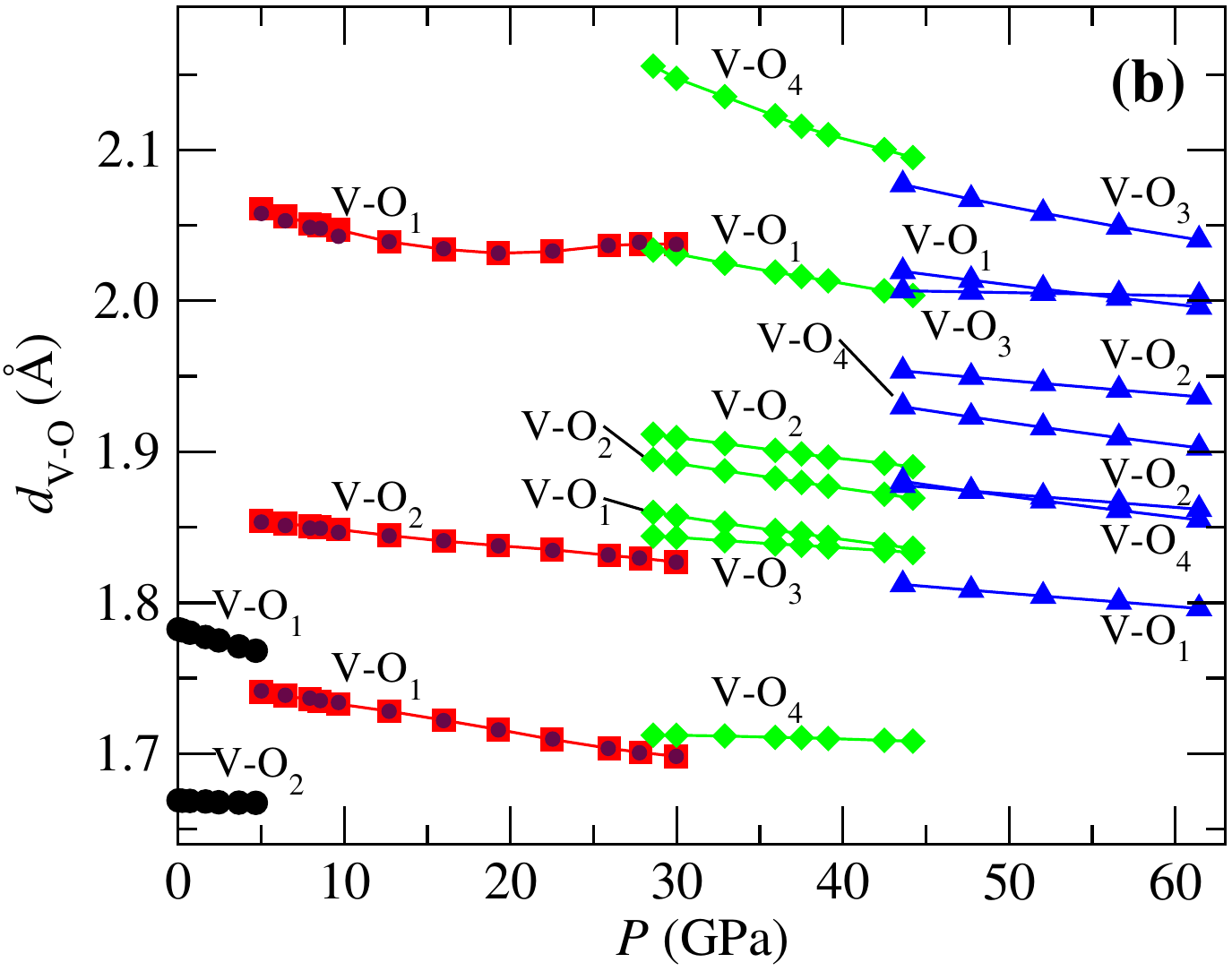}    
\end{tabular}
\caption{(Color online) Pressure dependence of interatomic bond distances (a) In-O 
and (b) V-O. Black circles, red squares, purple circles, green diamonds, and blue 
triangles correspond to CrVO$_4$-type, wolframite, CuWO$_4$-type, raspite, and 
AgMnO$_4$-type structures, respectively.}
\label{fig:6}
\end{figure}

According to our calculations, the CrVO$_4$-type structure is stable up to 4.4 GPa. 
For this phase there is a good agreement with the experimental data for the pressure 
evolution of volume and lattice parameters as is shown in Fig.~\ref{fig:4} and 
\ref{fig:5}, respectively. We can see from Fig.~\ref{fig:5} that lattice parameter $b$ 
is more compressible than $c$ and $a$. This arise from the fact that the InO$_6$ 
unit rotates around the $y$ axis as pressure increases because the apical interatomic 
bond distance of InO$_6$ polyhedra (In-O$_2$  in Fig.~\ref{fig:6}) is less 
compressible than equatorial ones (In-O$_1$); this promotes a shortening in the 
lattice parameter $b$. This behavior, also observed in CaSO$_4$~\cite{Gracia2012}, 
differs from that observed in TiSiO$_4$,~\cite{Gracia2009} InPO$_4$, and 
TiPO$_4$~\cite{Lopez2012b} where the apical In-O$_2$ bond distance remains 
almost constant under pressure. On the other hand, since V-O$_2$ distances are 
oriented in the $c$ direction, the compression of this axis is due to the reduction of 
the equatorial bond distances of InO$_6$, as V-O$_2$ remain almost constant under 
pressure.

At 4.4 GPa the CrVO$_4$-type structure undergoes a first order phase transition 
as can be seen in Fig.~\ref{fig:3}. Experimentally, prior to the transition to 
wolframite phase, a transition to a phase IV was observed. However it was not 
possible to give a description of this structure.~\cite{Errandonea2013} According 
to Fig.~\ref{fig:2} (a) scheelite-type and CuWO$_4$-type structure are competitive 
with wolframite. In order to find the phase IV we compare the simulated X-ray 
diffraction patterns of these structures with the experimental ones reported in 
Ref.~\citenum{Errandonea2013}, however the patterns of scheelite-type structure 
do not fit with the experimental one. We also compare the simulated X-ray patterns 
of the products of decomposition of InVO$_4$ and we did not find an agreement 
with the diffraction patterns of InO+VO$_3$. Hence, the description of phase IV 
will be left for a future work. According to our results, the wolframite and 
CuWO$_4$-type structure has almost the same energy, where the CuWO$_4$-type 
could be considered as a distortion of the wolframite; hence, the crystallographic 
parameters from the CuWO$_4$-type structure are almost the same as the wolframite, 
see Table~\ref{table:1} to \ref{table:3} and Figures~\ref{fig:2} to \ref{fig:6}. We will 
see in the next section that phonons from wolframite and the CuWO$_4$-type 
structure are very similar.

In the transition from CrVO$_4$-type to wolframite, InVO$_4$ has a large 
volume reduction of 17.5\%, in good agreement with the experimental value 
(see Figure~\ref{fig:4}). This volume reduction is associated with a change 
in the coordination of the polyhedron around V, from VO$_4$ to VO$_6$. 
Figure~\ref{fig:5} shows that the lattice parameters and $\beta$ angle 
obtained from calculations are in very good agreement with the experimental 
results. 
According to Figure~\ref{fig:5}b, the lattice parameter $b$ is much 
more compressible than $a$ and $c$, here $a$ almost remains constant. 
In-O bond lengths in this phase behave in a way similar to that for 
CrVO$_4$-type. 
The apical bond distance (In-O$_2$ in the middle zone of 
Figure~\ref{fig:6}a) is less compressible than the other In-O bond distances. 
As seen in Figure~\ref{fig:6}, a change in the slope of the pressure 
dependence of interatomic bond distances V-O$_1$ and In-O starts at 14 
GPa. 
These changes are not reflected in the lattice parameters but they play 
an important role in the vibrational properties that will be discussed in 
section~\ref{sIII-C}.

According to Fig.~\ref{fig:3} the wolframite phase is stable up to 28.1 GPa, then 
InVO$_4$ presents a transition to the raspite structure (phase VI). This transition 
is accompanied of a volume reduction of 7.5 \%. Experiments on InVO$_4$ were 
conducted only up to 24 GPa,~\cite{Errandonea2013} reason for which this phase 
transition was not observed. The lattice parameters of raspite for a volume of 212 
\AA$^3$ and a pressure of 32.9 GPa are $a$ = 4.9869 \AA, $b$ = 4.7032 \AA, $c$ 
= 9.0390 \AA, and $\beta$ = 90.51$^{\circ}$. In this structure all the non equivalent 
atoms are located in the 4$e$ WP: In (0.5334, 0.2408, 0.6182), V (0.03432, 0.2457, 
0.8893), O1 (0.2290, 0.9886, 0.03418), O2 (0.2089, 0.003844, 0.5248), O3 (0.3813, 
0.2795, 0.8235), and O4 (0.09154, 0.8490, 0.7789). We found from the EOS a bulk 
modulus of $B_0$ = 112.72 GPa with $V_0$ = 252.77 \AA$^3$ and $B_0$' = 4.25. 
From lattice parameters and $\beta$ angle presented in Fig.\ref{fig:5} we observe 
that this phase resembles a distorted tetragonal structure, it consists of alternating 
zig-zag chains of InO$_7$ polyhedra in the $z$ direction linked by VO$_7$ units. 
InO$_7$ polyhedra share edges in the zig-zag chains running perpendicular to plane 
[010], whereas they just share the corners in the $z$ direction. The same behavior 
was observed for VO$_7$ polyhedra. Figure~\ref{fig:6} shows that both In-O and 
V-O interatomic bond distances have similar compressibility, except for one V-O$_4$ 
and one In-O$_2$ bond distances that are almost constants in all the range of 
pressure stability.

As pressure increases, respite becomes unstable and InVO$_4$ undergoes a phase 
transition to the AgMnO$_4$-type structure at 44 GPa (phase VII). As can be seen 
in Fig.~\ref{fig:4} there is just a small volume reduction in the transition from 
raspite to the AgMnO$_4$-type structure, $\Delta V$ = -2.9 \%. At 52 GPa the 
lattice parameters of this phase are $a$ = 5.3266 \AA, $b$ = 5.1199 \AA, $c$ = 
8.4399 \AA, and $\beta$ = 122.56 $^{\circ}$, the volume 194 \AA$^3$. The 
Figure~\ref{fig:5} shows that although raspite and AgMnO$_4$-type phases 
belongs to the same SG, they have marked differences in the lattice parameters, 
however they have almost the same compressibility. This is also reflected in the 
change of interatomic bond distances as function of pressure. Like in raspite, in this 
phase the In, V, and O atoms are located in the 4$e$ WP as follow: In (0.3725, 
0.06524, 0.12899), V (0.9727, 0.005625, 0.6706), O1 (0.7477, 0.9612, 0.4191), O2 
(0.9497, 0.3533, 0.5960), O3 (0.3198, 0.2113, 0.8709), and O4 (0.7676, 0.2088, 
0.7487). This phase is formed by InO$_8$ and VO$_8$ units that shares edges and 
corners. The InO$_8$ polyhedra forms layers that lay in the $bc$ plane, separated 
by layers formed by VO$_8$ polyhedra. We found that InVO$_4$ is stable in this 
phase up to 62 GPa, which is the highest pressure reached in this study.

\subsection{Vibrational properties}\label{sIII-C} 

Nowadays the lattice vibration studies, through the analysis of Raman (R) and 
Infrared (IR) spectra from experimental as well as theoretical methods, have 
become fundamental tools to understand the behavior of materials at ambient 
conditions and under extreme conditions of temperature and 
pressure.~\cite{Muj03,Baroni2001} In particular, these studies help to determine 
and realize whether a phase transition takes place. In many cases the experimental 
and theoretical approximations are conjugated with great success to study 
$A$VO$_4$ compounds, see for example references \citenum{Panchal2011}, 
\citenum{Panchal2011c}, \citenum{Errandonea2013b}, \citenum{Errandonea2013c} 
and references there in; whereas in other circumstances the theoretical results serve 
as a guide for future experimental and theoretical studies.~\cite{Lopez2012b,
Lopez2015} In this section we analyze the lattice dynamics of InVO$_4$ in the 
phases III, V, raspite (VI) and AgMnO$_4$-type (VII) by means of the calculated 
phonon frequencies at $\Gamma$ point for each phase, their pressure dependence, 
and Gr\"uneisen parameters as well as their dispersion relation along the Brillouin 
zone and phonon DOS.

\setlength{\tabcolsep}{1.8pt}
\begin{table*}[ht!]
\caption
{Calculated Raman and infrared phonon frequencies for CrVO$_4$-type phase (at 
ambient pressure) and wolframite phase (at 6.44 GPa) of InVO$_4$ at the 
$\Gamma$ point. Frequencies $\omega$ are in cm$^{-1}$ and $d\omega/dP$ in 
cm$^{-1}$/GPa.}
\begin{tabular}{lccccclccccclccc}
\hline
&\multicolumn{5}{c}{CrVO$_4$-type} & & \multicolumn{5}{c}{wolframite} & & \multicolumn{3}{c}{CuWO$_4$-type}    \\
&\multicolumn{3}{c}{DFT} & \multicolumn{2}{c}{Exp.~\cite{Errandonea2013}} & & \multicolumn{3}{c}{DFT} & \multicolumn{2}{c}{Exp.~\cite{Errandonea2013}} & &
\multicolumn{3}{c}{DFT} \\
\cline{2-4} \cline{5-6} \cline{8-10} \cline{11-12} \cline{14-16}     
& $\omega$ & $d\omega/dp$ & $\gamma$ & $\omega$ & $d\omega/dp$ & & $\omega$ & $d\omega/dp$ & $\gamma$ & $\omega$ & $d\omega/dp$ & & $\omega$ & 
$d\omega/dp$ & $\gamma$ \\
\hline
$T(B_{3g})$    & 128.36 & 1.47 & 0.37 & 135 & 2.1 & $B_g$ & 108.15 & 0.13 & 0.24 & 109 & 1.8 & $A_g$ & 108.15 & 0.12 & 0.22  \\   
$T(B_{1g})$    & 153.78 &-2.38 &-0.60 & 191 & 0.7 & $A_g$ & 124.43 & 0.05 & 0.08 & 118 & 1.9 & $A_g$ & 124.39 & 0.05 & 0.08  \\  
$T(A_g)$       & 193.24 & 0.27 & 0.07 & 218 & 4.5 & $B_g$ & 139.17 & 0.75 & 1.04 & 145 & 2.0 & $A_g$ & 139.24 & 0.76 & 1.05  \\     
$R(B_{1g})$    & 208.32 & 4.93 & 1.24 & 252 & 3.6 & $B_g$ & 189.11 & 1.42 & 1.44 & 149 &-0.1 & $A_g$ & 189.01 & 1.43 & 1.45  \\  
$R(B_{2g})$    & 237.68 & 3.88 & 0.98 & 342 & 0.4 & $B_g$ & 214.33 & 1.74 & 1.55 & 204 & 1.5 & $A_g$ & 214.26 & 1.77 & 1.57  \\    
$\nu_2(A_g)$   & 334.48 & 0.37 & 0.09 & 348 & 5.6 & $A_g$ & 241.95 &-0.52 &-0.43 & 223 & 2.2 & $A_g$ & 242.18 &-0.58 &-0.48  \\
$R(B_{3g})$    & 352.16 & 5.01 & 1.27 & 377 & 1.9 & $B_g$ & 289.68 & 1.06 & 0.71 & 241 & 0.0 & $A_g$ & 289.61 & 1.07 & 0.72  \\   
$\nu_4(B_{1g})$& 361.53 & 1.38 & 0.35 & 389 & 4.4 & $A_g$ & 304.06 & 0.74 & 0.47 & 251 & 1.1 & $A_g$ & 304.26 & 0.72 & 0.46  \\
$\nu_2(B_{2g})$& 370.74 & 4.64 & 1.17 & 390 & 1.4 & $A_g$ & 340.75 & 1.23 & 0.70 & 319 & 1.2 & $A_g$ & 340.99 & 1.13 & 0.65  \\
$\nu_4(A_g)$   & 380.68 & 1.33 & 0.34 & 456 & 5.2 & $B_g$ & 379.81 & 1.95 & 1.00 & 336 & 2.4 & $A_g$ & 379.61 & 1.95 & 0.99  \\
$\nu_4(B_{3g})$& 422.85 & 6.14 & 1.55 & 637 & 7.2 & $A_g$ & 420.14 & 1.90 & 0.88 & 347 & 1.6 & $A_g$ & 420.11 & 1.88 & 0.87  \\
$\nu_3(B_{1g})$& 657.99 & 7.59 & 1.92 & 755 & 5.7 & $B_g$ & 440.53 & 3.87 & 1.67 & 378 & 1.9 & $A_g$ & 440.39 & 3.87 & 1.67  \\  
$\nu_3(A_g)$   & 752.49 & 6.36 & 1.61 & 847 & 4.2 & $B_g$ & 499.37 & 3.37 & 1.30 & 433 & 2.0 & $A_g$ & 499.34 & 3.39 & 1.30  \\
$\nu_1(A_g$)   & 920.05 & 1.33 & 0.34 & 914 & 1.3 & $A_g$ & 520.75 & 2.32 & 0.86 & 531 & 1.8 & $A_g$ & 520.79 & 2.29 & 0.85  \\
$\nu_3(B_{3g})$& 925.12 & 2.34 & 0.59 & 918 & 2.1 & $B_g$ & 679.84 & 5.04 & 1.41 & 684 & 5.1 & $A_g$ & 679.70 & 5.13 & 1.44  \\
               &        &      &      &     &     & $A_g$ & 712.73 & 4.22 & 1.14 & 723 & 3.8 & $A_g$ & 712.93 & 4.20 & 1.13  \\
               &        &      &      &     &     & $B_g$ & 740.12 & 5.15 & 1.33 & 758 & 5.2 & $A_g$ & 740.05 & 5.20 & 1.34  \\
               &        &      &      &     &     & $A_g$ & 835.15 & 4.62 & 1.07 & 850 & 4.4 & $A_g$ & 836.02 & 4.62 & 1.06  \\
\\
$T(B_{1u})$     & 102.64 & 0.14 & 0.04 &  &  & $B_u$ &  72.15 &-7.61 &-11.97 &  &  & $A_u$ &  72.49 &-7.98 &-14.00  \\ 
$T(B_{3u})$     & 150.91 &-2.09 &-0.53 &  &  & $A_u$ & 175.83 &-0.08 & -0.09 &  &  & $A_u$ & 176.03 &-0.17 & -0.20  \\   
$T(B_{1u})$     & 152.31 & 0.85 & 0.22 &  &  & $B_u$ & 200.95 &-0.05 & -0.05 &  &  & $A_u$ & 201.02 &-0.08 & -0.07  \\   
$T(B_{2u})$     & 214.39 & 0.20 & 0.05 &  &  & $B_u$ & 267.83 &-0.43 & -0.32 &  &  & $A_u$ & 267.80 &-0.45 & -0.33  \\   
$R(B_{3u})$     & 253.72 & 4.42 & 1.13 &  &  & $A_u$ & 293.65 &-0.34 & -0.23 &  &  & $A_u$ & 293.68 &-0.38 & -0.25  \\   
$T(B_{2u})$     & 253.85 & 1.21 & 0.31 &  &  & $B_u$ & 316.80 & 2.58 &  1.55 &  &  & $A_u$ & 316.73 & 2.58 &  1.55  \\   
$\nu_4(B_{3u})$ & 328.31 & 2.06 & 0.52 &  &  & $A_u$ & 361.33 &-0.60 & -0.33 &  &  & $A_u$ & 361.33 &-0.61 & -0.33  \\   
$R(B_{1u})$     & 346.36 & 6.73 & 1.71 &  &  & $B_u$ & 366.10 & 1.75 &  0.93 &  &  & $A_u$ & 365.90 & 1.74 &  0.92  \\   
$\nu_2(B_{2u}$) & 358.37 &-0.03 &-0.01 &  &  & $A_u$ & 456.57 & 3.04 &  1.28 &  &  & $A_u$ & 456.67 & 3.01 &  1.26  \\   
$\nu_4(B_{2u}$) & 396.86 & 3.69 & 0.94 &  &  & $B_u$ & 497.00 & 3.84 &  1.48 &  &  & $A_u$ & 496.97 & 3.85 &  1.48  \\   
$\nu_4(B_{1u}$) & 423.21 & 4.95 & 1.26 &  &  & $B_u$ & 531.83 & 4.27 &  1.53 &  &  & $A_u$ & 531.79 & 4.31 &  1.54  \\   
$\nu_3(B_{3u}$) & 681.81 & 7.29 & 1.86 &  &  & $A_u$ & 544.80 & 3.75 &  1.32 &  &  & $A_u$ & 544.90 & 3.74 &  1.31  \\   
$\nu_1(B_{2u}$) & 758.20 & 6.59 & 1.68 &  &  & $A_u$ & 610.39 & 4.98 &  1.56 &  &  & $A_u$ & 610.49 & 5.04 &  1.57  \\   
$\nu_3(B_{1u}$) & 893.20 & 1.72 & 0.44 &  &  & $B_u$ & 697.88 & 5.14 &  1.41 &  &  & $A_u$ & 698.08 & 5.27 &  1.44  \\   
$\nu_3(B_{2u}$) & 936.59 & 0.94 & 0.24 &  &  & $A_u$ & 761.40 & 3.91 &  1.00 &  &  & $A_u$ & 760.86 & 3.95 &  1.00   \\
\hline
\end{tabular}
\label{table:4}
\end{table*}

\setlength{\tabcolsep}{4pt}
\begin{table}[ht!]
\caption
{Calculated Raman frequencies for raspite (at 30 GPa) and AgMnO$_4$-type structure 
(at 47.7 GPa) of InVO$_4$ at the $\Gamma$ point. Frequencies $\omega$ are in 
cm$^{-1}$ and $d\omega/dP$ in cm$^{-1}$/GPa.}
\begin{tabular}{lccclccc}
\hline
& \multicolumn{3}{c}{raspite} & & \multicolumn{3}{c}{AgMnO$_4$-type} \\ 
\cline{2-4} \cline{6-8} 
& $\omega$ & $d\omega/dp$ & $\gamma$ & & $\omega$ & $d\omega/dp$ & $\gamma$  \\
\hline            
$B_g$ &  149.38 & 0.02 & 0.03 & $B_g$ & 141.70 & 0.26 & 0.60   \\  
$A_g$ &  154.28 & 0.55 & 0.88 & $A_g$ & 153.75 & 0.51 & 1.15   \\   
$B_g$ &  160.19 & 0.89 & 1.41 & $B_g$ & 172.96 & 0.43 & 0.98   \\   
$A_g$ &  160.95 & 0.68 & 1.08 & $A_g$ & 189.07 & 0.60 & 1.36   \\   
$B_g$ &  175.50 & 0.51 & 0.81 & $A_g$ & 199.15 & 0.36 & 0.82   \\   
$A_g$ &  188.77 & 0.64 & 1.01 & $B_g$ & 224.80 & 0.69 & 1.56   \\   
$A_g$ &  196.31 & 1.51 & 2.39 & $A_g$ & 241.85 & 0.98 & 2.23   \\   
$A_g$ &  230.57 & 1.69 & 2.67 & $B_g$ & 308.66 & 0.83 & 1.88   \\   
$B_g$ &  268.70 & 1.31 & 2.08 & $A_g$ & 312.23 & 0.61 & 1.40   \\   
$B_g$ &  272.34 & 1.50 & 2.37 & $A_g$ & 332.98 & 1.34 & 3.06   \\   
$A_g$ &  287.58 & 1.36 & 2.15 & $B_g$ & 349.16 & 1.05 & 2.40   \\   
$A_g$ &  306.89 & 1.82 & 2.87 & $A_g$ & 362.50 & 1.29 & 2.95   \\   
$B_g$ &  313.30 & 0.81 & 1.28 & $A_g$ & 383.38 & 1.31 & 2.98   \\   
$B_g$ &  317.27 & 1.79 & 2.83 & $B_g$ & 388.12 & 1.35 & 3.08   \\   
$A_g$ &  366.91 & 1.39 & 2.20 & $A_g$ & 430.25 & 1.32 & 3.00   \\   
$A_g$ &  376.11 & 1.58 & 2.51 & $B_g$ & 434.29 & 1.33 & 3.04   \\   
$B_g$ &  382.82 & 1.48 & 2.34 & $A_g$ & 465.14 & 1.11 & 2.53   \\   
$B_g$ &  395.89 & 0.97 & 1.54 & $B_g$ & 471.55 & 1.36 & 3.09   \\   
$A_g$ &  399.66 & 0.63 & 1.00 & $B_g$ & 474.75 & 1.50 & 3.42   \\   
$B_g$ &  404.40 & 1.80 & 2.84 & $A_g$ & 485.99 & 1.78 & 4.05   \\   
$A_g$ &  448.47 & 1.51 & 2.39 & $B_g$ & 502.54 & 1.45 & 3.31   \\   
$B_g$ &  472.82 & 1.74 & 2.75 & $B_g$ & 530.49 & 1.65 & 3.76   \\   
$A_g$ &  481.82 & 1.56 & 2.46 & $A_g$ & 555.38 & 1.93 & 4.40   \\   
$B_g$ &  503.14 & 1.76 & 2.78 & $A_g$ & 563.88 & 2.49 & 5.68   \\   
$A_g$ &  519.02 & 1.80 & 2.85 & $B_g$ & 568.75 & 1.94 & 4.43   \\   
$B_g$ &  549.77 & 2.26 & 3.58 & $B_g$ & 593.57 & 1.60 & 3.64   \\   
$B_g$ &  630.57 & 2.60 & 4.12 & $B_g$ & 636.40 & 1.94 & 4.41   \\   
$A_g$ &  638.81 & 2.73 & 4.33 & $A_g$ & 640.11 & 1.84 & 4.19   \\   
$B_g$ &  683.74 & 2.26 & 3.58 & $A_g$ & 695.62 & 2.25 & 5.12   \\   
$A_g$ &  729.24 & 2.45 & 3.87 & $B_g$ & 719.83 & 2.02 & 4.61   \\   
$B_g$ &  729.71 & 2.45 & 3.87 & $B_g$ & 724.57 & 2.46 & 5.62   \\   
$A_g$ &  750.96 & 2.51 & 3.98 & $A_g$ & 738.48 & 2.32 & 5.28   \\   
$A_g$ &  784.38 & 1.91 & 3.02 & $B_g$ & 773.84 & 2.43 & 5.53   \\   
$A_g$ &  817.94 & 2.78 & 4.41 & $A_g$ & 778.88 & 2.33 & 5.30   \\   
$B_g$ &  821.21 & 3.05 & 4.83 & $B_g$ & 825.34 & 2.41 & 5.50   \\   
$B_g$ &  902.50 & 2.00 & 3.17 & $A_g$ & 863.67 & 2.39 & 5.46   \\
\hline
\end{tabular}
\label{table:5}
\end{table}

\setlength{\tabcolsep}{4pt}
\begin{table}[ht!]
\caption
{Calculated infrared frequencies for raspite (at 30 GPa) and AgMnO$_4$-type 
structure (at 47.7 GPa) of InVO$_4$ at the $\Gamma$ point. Frequencies 
$\omega$ are in cm$^{-1}$ and $d\omega/dP$ in cm$^{-1}$/GPa.}
\begin{tabular}{lccclccc}
\hline
& \multicolumn{3}{c}{raspite} & & 
\multicolumn{3}{c}{AgMnO$_4$-type} \\ 
\cline{2-4} \cline{6-8} 
& $\omega$ & $d\omega/dp$ & $\gamma$ & & $\omega$ & 
$d\omega/dp$ & $\gamma$  \\
\hline            
$B_u$ &   80.13 & 0.57 & 0.91 & $A_u$ & 159.25 & 0.67 & 1.53     \\  
$A_u$ &  106.41 & 0.84 & 1.34 & $A_u$ & 193.84 & 0.58 & 1.33     \\   
$A_u$ &  119.36 & 0.82 & 1.29 & $B_u$ & 195.88 & 0.61 & 1.40     \\   
$B_u$ &  176.56 & 1.14 & 1.81 & $B_u$ & 249.45 & 1.16 & 2.66     \\   
$A_u$ &  183.84 & 0.69 & 1.09 & $A_u$ & 253.19 & 0.94 & 2.16     \\   
$B_u$ &  225.00 & 0.56 & 0.89 & $A_u$ & 278.71 & 1.05 & 2.40     \\   
$A_u$ &  243.98 & 1.27 & 2.01 & $B_u$ & 289.05 & 1.21 & 2.77     \\   
$A_u$ &  260.53 & 2.42 & 3.83 & $A_u$ & 333.88 & 0.98 & 2.24     \\   
$B_u$ &  289.55 & 1.65 & 2.62 & $A_u$ & 361.07 & 1.13 & 2.59     \\   
$A_u$ &  289.91 & 1.05 & 1.67 & $B_u$ & 366.60 & 1.62 & 3.70     \\   
$B_u$ &  312.00 & 1.59 & 2.51 & $A_u$ & 387.75 & 0.89 & 2.05     \\   
$A_u$ &  330.01 & 1.38 & 2.19 & $B_u$ & 393.32 & 1.40 & 3.21     \\   
$B_u$ &  339.55 & 2.37 & 3.75 & $A_u$ & 424.11 & 1.71 & 3.90     \\   
$A_u$ &  357.23 & 1.94 & 3.08 & $B_u$ & 435.82 & 1.01 & 2.32     \\   
$B_u$ &  379.75 & 0.88 & 1.40 & $A_u$ & 451.90 & 2.28 & 5.22     \\   
$A_u$ &  401.66 & 1.68 & 2.66 & $B_u$ & 453.84 & 1.59 & 3.63     \\   
$B_u$ &  420.91 & 2.45 & 3.89 & $B_u$ & 464.04 & 1.91 & 4.37     \\   
$A_u$ &  443.83 & 1.88 & 2.98 & $A_u$ & 476.15 & 1.29 & 2.94     \\   
$B_u$ &  455.67 & 2.06 & 3.26 & $B_u$ & 505.51 & 1.40 & 3.20     \\   
$A_u$ &  491.46 & 2.09 & 3.31 & $A_u$ & 508.74 & 2.16 & 4.93     \\   
$B_u$ &  493.27 & 2.02 & 3.20 & $B_u$ & 560.65 & 1.58 & 3.60     \\   
$B_u$ &  522.22 & 2.75 & 4.35 & $B_u$ & 601.75 & 1.81 & 4.15     \\   
$A_u$ &  526.72 & 2.15 & 3.41 & $A_u$ & 623.33 & 1.71 & 3.91     \\   
$B_u$ &  548.07 & 1.93 & 3.05 & $B_u$ & 629.23 & 1.96 & 4.48     \\   
$A_u$ &  555.61 & 2.01 & 3.19 & $A_u$ & 650.58 & 2.29 & 5.23     \\   
$A_u$ &  605.22 & 2.39 & 3.78 & $B_u$ & 659.22 & 2.11 & 4.82     \\   
$B_u$ &  606.18 & 2.22 & 3.51 & $A_u$ & 680.57 & 1.91 & 4.36     \\   
$B_u$ &  676.83 & 2.50 & 3.96 & $B_u$ & 724.84 & 2.36 & 5.38     \\   
$A_u$ &  701.39 & 2.73 & 4.33 & $A_u$ & 732.64 & 2.38 & 5.43     \\   
$B_u$ &  765.13 & 2.73 & 4.33 & $B_u$ & 776.58 & 2.19 & 4.99     \\   
$A_u$ &  790.99 & 2.16 & 3.43 & $A_u$ & 795.32 & 2.39 & 5.47     \\   
$A_u$ &  840.92 & 2.64 & 4.18 & $A_u$ & 813.90 & 2.23 & 5.10     \\   
$B_u$ &  850.70 & 2.45 & 3.87 & $B_u$ & 815.64 & 2.60 & 5.93     \\
\hline
\end{tabular}
\label{table:6}
\end{table}

\begin{figure*}[ht!]
\centering
\begin{tabular}{cc}
\includegraphics[width=8.5cm]{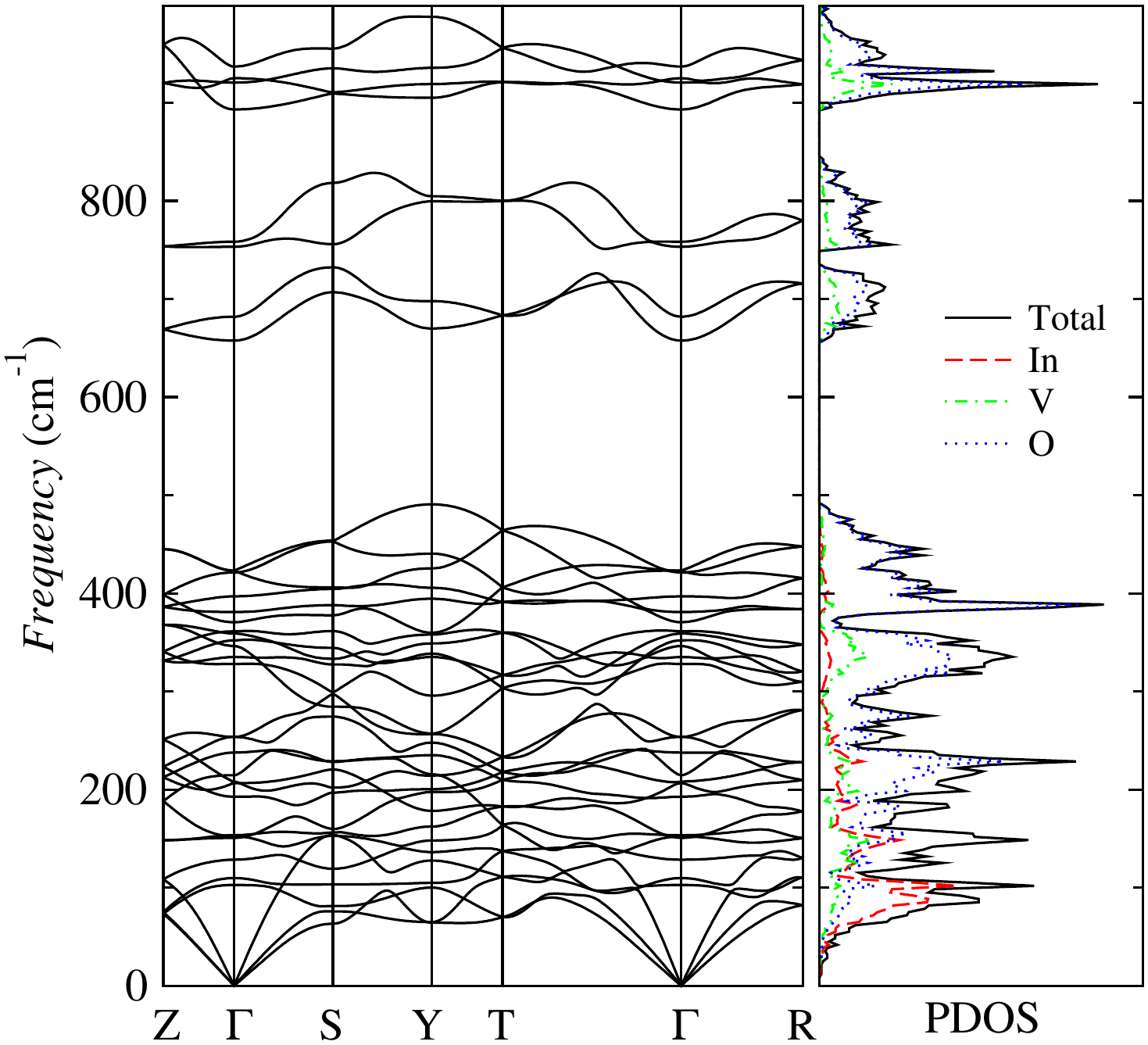}  &
\includegraphics[width=8.5cm]{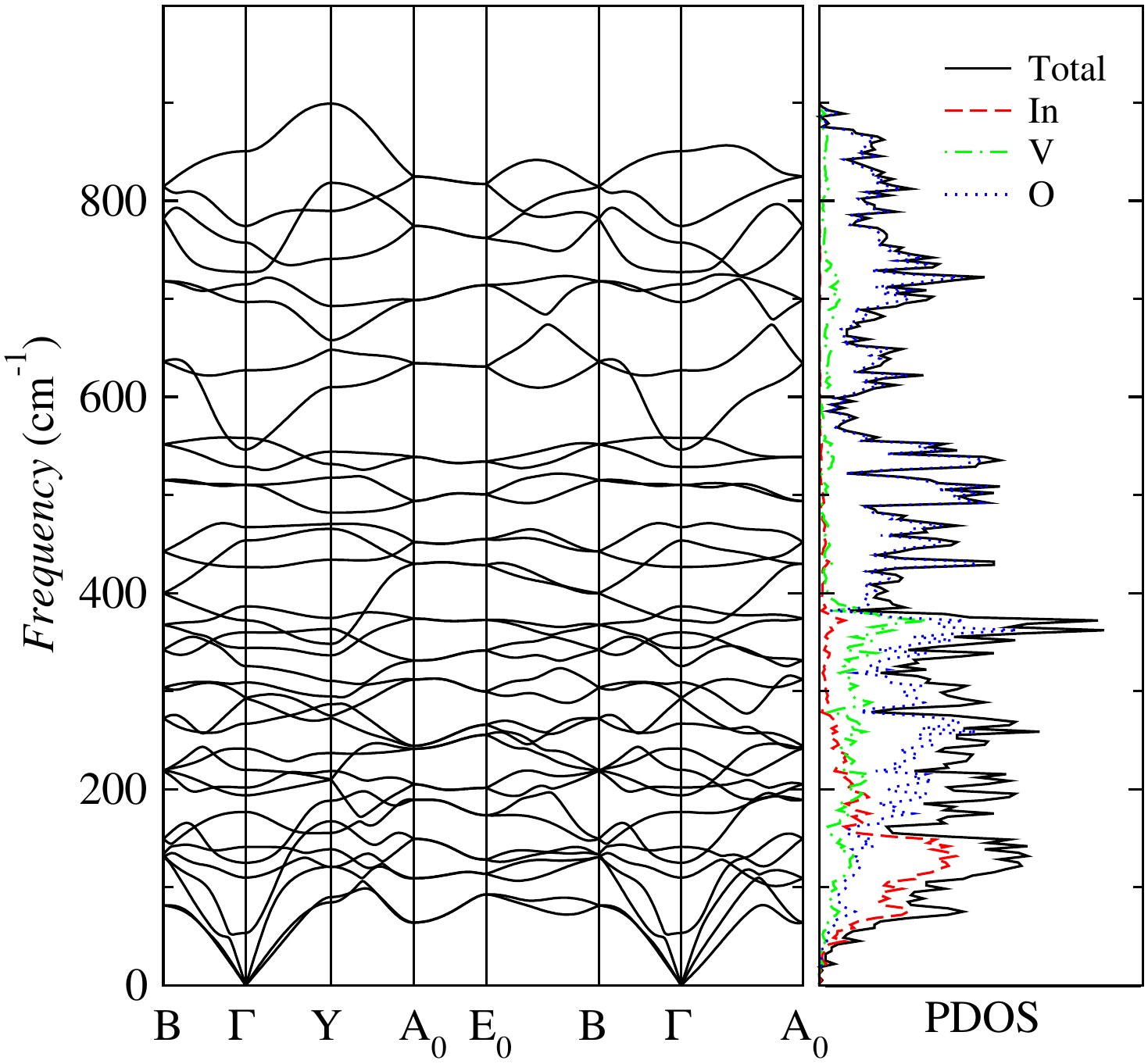} \\
(a) CrVO$_4$-type (at $\approx$ 0 GPa) & (b) wolframite  (at $\approx$ 9 GPa) \\
\\
\includegraphics[width=8.5cm]{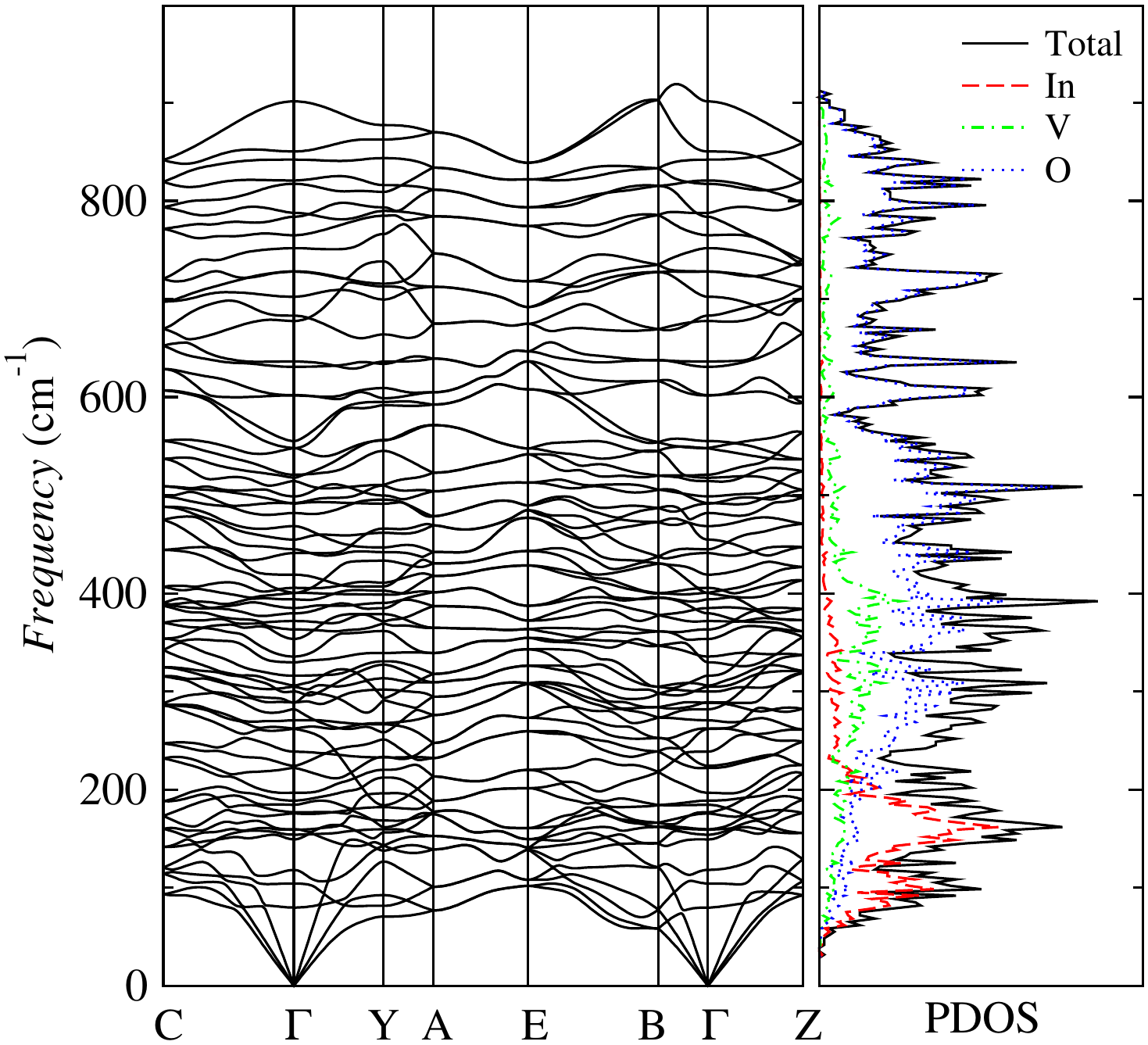} &
\includegraphics[width=8.5cm]{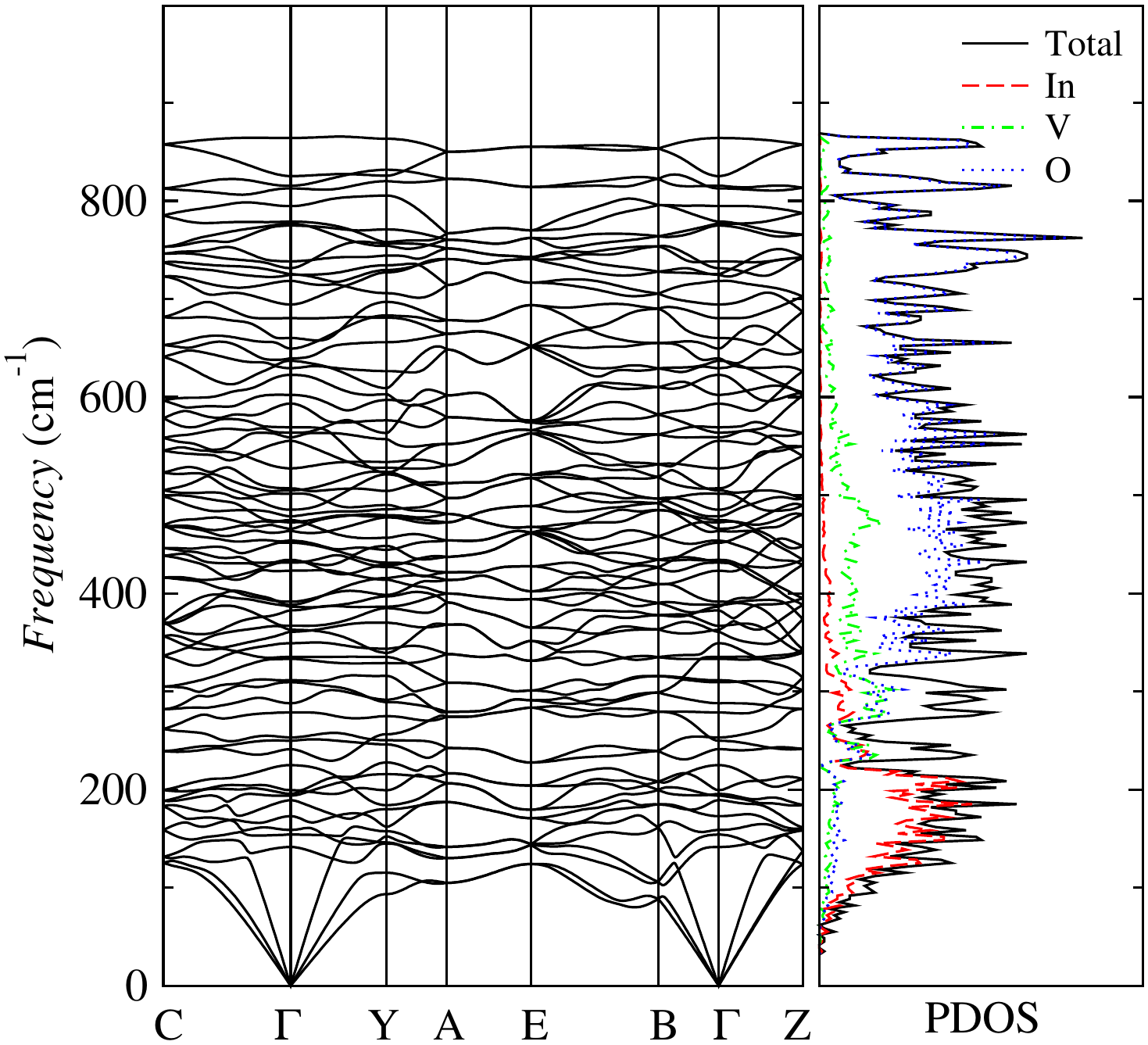} \\
(c) raspite (at 30 GPa) & (d) AgMnO$_4$-type (at 47.7 GPa)
\end{tabular}
\caption{(Color online) Phonon spectrum and phonon DOS of InVO$_4$ 
polymorphs: (a) CrVO$_4$-type (at $\approx$ 0 GPa), (b) wolframite (at 
$\approx$ 9 GPa), (c) raspite (at 30 GPa), and (d) AgMnO$_4$-type (at 47.7 GPa).}
\label{fig:7}
\end{figure*}

\begin{figure}[ht!]
\centering
\begin{tabular}{c}
\includegraphics[width=8.5cm]{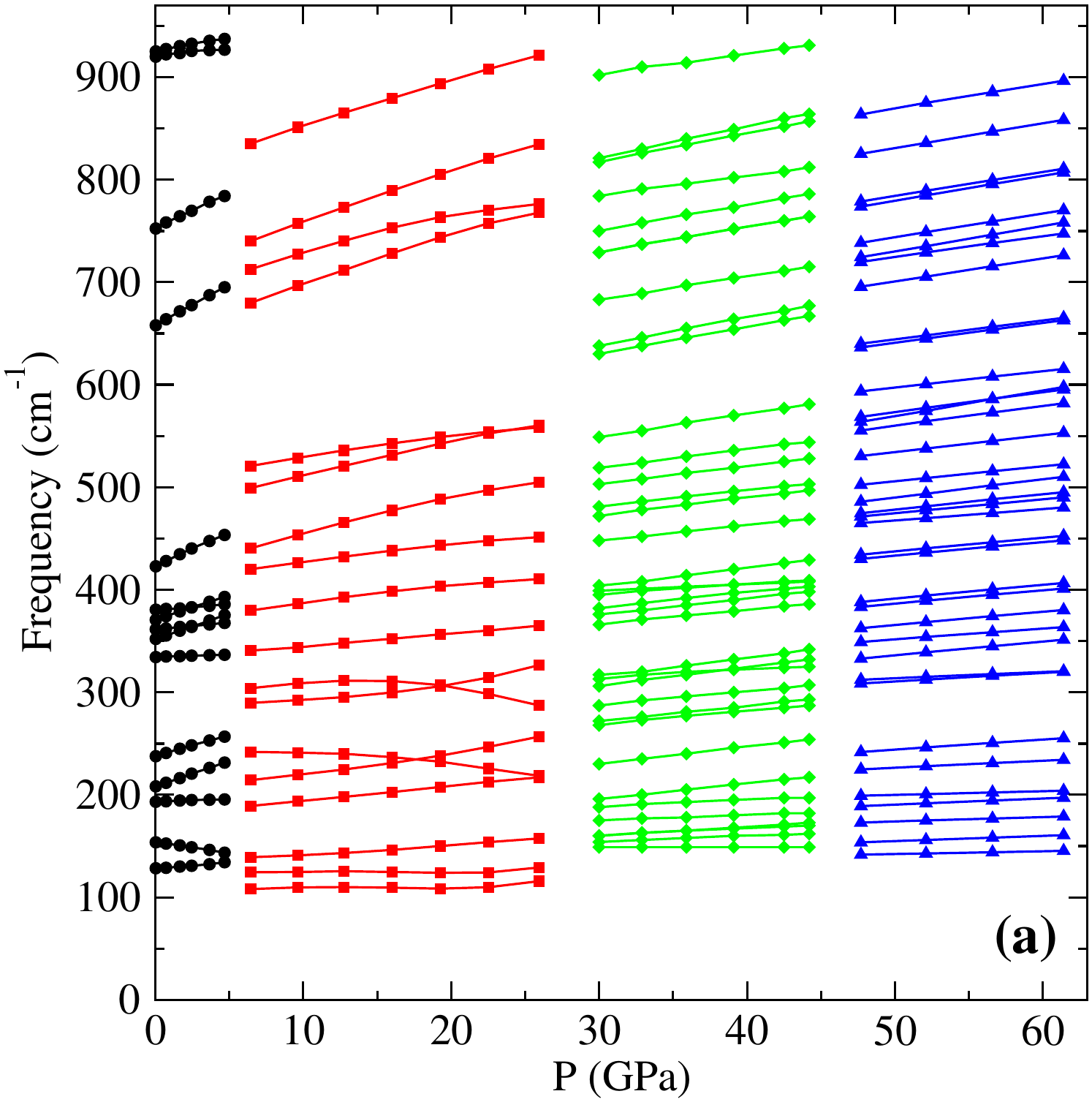}  \\
\includegraphics[width=8.5cm]{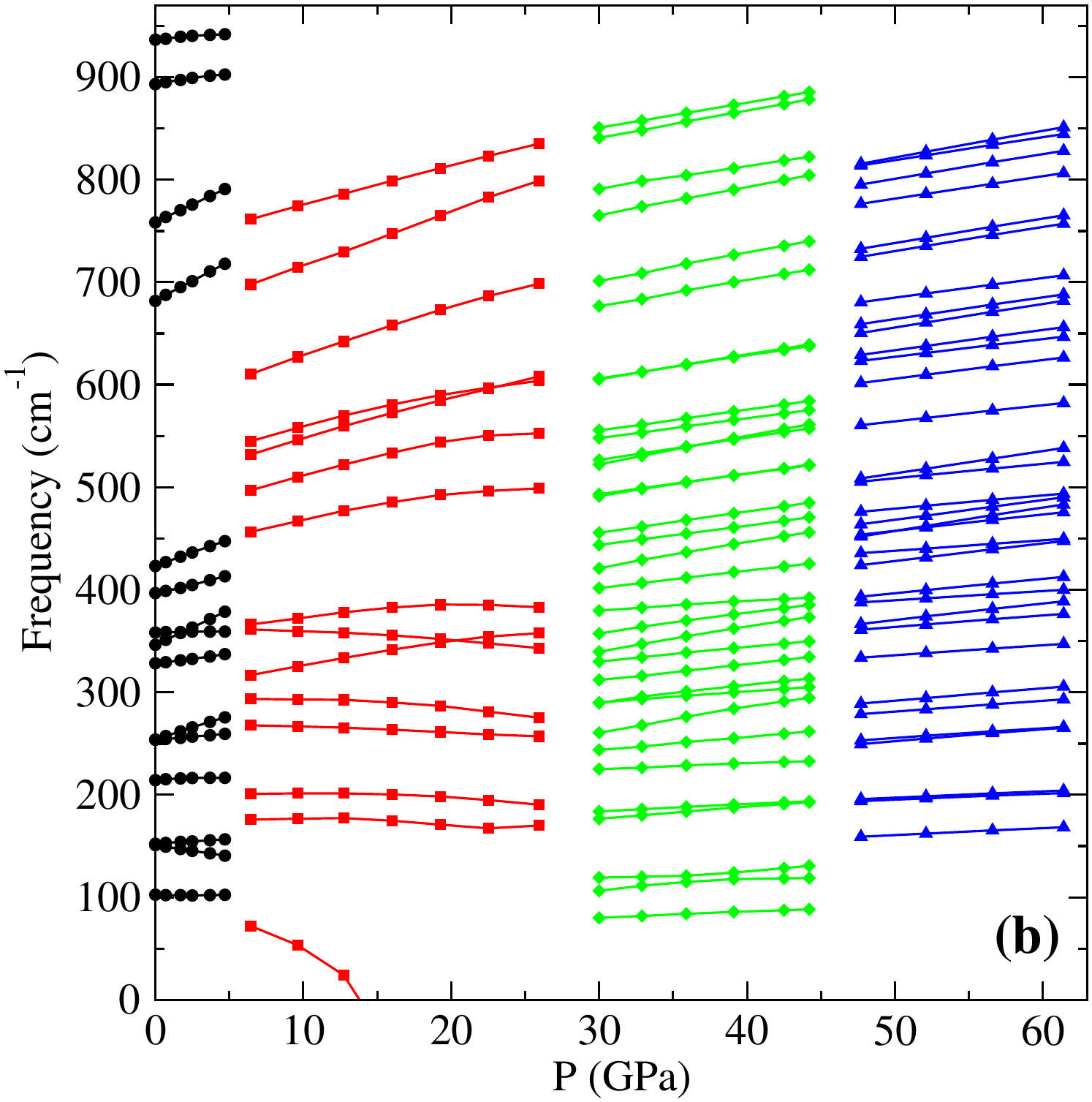}  
\end{tabular}
\caption{(Color online) Pressure dependence of (a) Raman and (b) IR frequencies 
of InVO$_4$. Black circles, red squares, green diamonds, and blue triangles 
corresponds to CrVO$_4$-type, wolframite, raspite and AgMnO$_4$-type structures,
respectively.}
\label{fig:8}
\end{figure}

The calculated and experimental R and IR frequencies at $\Gamma$ point for the 
studied phases appear in Table~\ref{table:4} to \ref{table:6}, the pressure 
coefficients ($d\omega/dP$,) and Gr\"uneisen parameters [$\gamma$ = -$\partial$
(ln $\omega$)/$\partial$(ln $V$)] were also included. The phonon relation of 
dispersion and the  phonon DOS are displayed in Fig.~\ref{fig:7}. The vibrational 
modes of InVO$_4$ can be classified as internal or external modes of the VO$_4$ 
unit. The external modes correspond either to a pure translation ($T$) or to a pure 
rotation ($R$) of the VO$_4$ unit. Whereas the internal modes of the VO$_4$ 
tetrahedra are $\nu_1$ (symmetric stretching), $\nu_2$ (symmetric bending), 
$\nu_3$ (asymmetric stretching), and $\nu_4$ 
(asymmetric bending).~\cite{Lopez2012b} The $T$ modes are usually the lowest in
frequency, the $\nu_x$ modes are the highest in frequency, and the frequencies of 
the $R$ modes are between those of the $T$ and $\nu_x$ modes.

According to the group theory analysis, the $Cmcm$ space group of CrVO$_4$-type 
structure has the following representation at $\Gamma$ point: $\Gamma$ = 5$A_g$ 
+ 4$B_{1g}$ + 6$B_{1u}$ + 3$A_u$ + 2$B_{2g}$ + 7$B_{2u}$ + 4$B_{3g}$ + 
5$B_{3u}$. Where there are three acoustic modes: $B_{1u}$, $B_{2u}$, and 
$B_{3u}$, three silent modes $A_u$, 15 infrared active modes: 5$B_{1u}$, 
6$B_{2u}$, and 4$B_{3u}$, and 15 Raman active modes: 5$A_g$, 4$B_{1g}$, 
2$B_{2g}$, and 4$B_{3g}$.~\cite{Lopez2012b} As can be seen in 
Table~\ref{table:4} this phase presents two $T$ and one $\nu_2$ phonons that are
characterized by a decrease of the vibrational frequency with pressure, i.e. negative 
pressure coefficients and Gr\"uneisen parameters, the Raman mode $B_{1g}$ 
(153.78 cm$^{-1}$) and two IR modes: $B_{3u}$ (150.91 cm$^{-1}$) and 
$B_{2u}$ (358.37 cm$^{-1}$). The softening of this $B_{1g}$ Raman ($B_{3u}$ 
IR) mode was also observed in CaSO$_4$~\cite{Gracia2012} 
(InPO$_4$),~\cite{Lopez2012b} and could be related with the instability of the 
CrVO$_4$-type structure under pressure. The pressure evolution of Raman and IR 
modes of the studied phases is displayed in Fig.~\ref{fig:8} (a) and (b), respectively. 
The phonon dispersion for CrVO$_4$, Fig.~\ref{fig:7} (a), shows no imaginary 
frequency. We also realized simulation of phonon dispersions up to $\approx$7 GPa 
and no significant changes were observed. As seen on the phonon DOS, phonons 
above 600 cm$^{-1}$ belong to internal $\nu_x$ modes, i.e. there are only 
contributions from V and O. The frequencies of the external modes are located 
below 255 cm$^{-1}$, and the intermediate is occupied by external ($R$) and 
internal ($\nu$) modes.

For wolframite the group theory predicts the following $\Gamma$ phonon modes: 
$\Gamma$ = 8$A_g$ + 10$B_g$ + 8$A_u$ + 10$B_u$. Here 2$B_u$ and one 
$A_u$ infrared modes are acoustic, 18 are Raman ($8A_g + 10B_g$) and 15 are 
infrared ($7A_u + 9B_u$). The group theory predicts the same modes for 
CuWO$_4$-type structure at $\Gamma$ point. According to Table~\ref{table:4} 
both phases wolframite and CuWO$_4$-type structure presents almost the same 
phonon frequencies at 6.44 GPa. Besides, we observed that phonon spectrum for 
both phases are very similar, reason for which we only include the spectrum from 
wolframite.

Figure~\ref{fig:8} (a) shows that the slope of some Raman modes of wolframite 
starts to change around 16 GPa, so that we calculated the pressure coefficients 
and Gr\"uneisen parameters from Table~\ref{table:4} by considering just the 
frequencies from 6 to 14 GPa which correspond to the linear trend for wolframite 
phase. This change in the frequencies of wolframite is due to a shift in the slope of 
the In-O$_x$ and V-O$_1$ interatomic bond distances as pressure increases, see 
Fig.~\ref{fig:6}. Having in mind this, we see that wolframite presents several R 
and IR phonon modes with negative pressure coefficients. It is noteworthy that 
$B_u$ IR mode softens completely around 14 GPa, to our knowledge this behavior 
was not observed previously in other CrVO$_4$-type compounds. However, it has 
been also reported that this mode softens in the wolframite high-pressure phase of 
InPO$_4$ and TiPO$_4$,~\cite{Lopez2012b} and in other compounds that 
crystallizes at ambient pressure in the wolframite structure such as 
MgWO$_4$~\cite{Ruiz2010} and InTaO$_4$.~\cite{Errandonea2016} The 
Figure~\ref{fig:7} (b) shows the phonon spectra and phonon DOS of wolframite at 
$\approx$9 GPa. We have to mention that we had to perform more calculations 
of the phonon spectrum with bigger supercells in specific directions of the BZ in 
order to eliminate possible errors from the supercell method with the PHONON
program. On the other hand, above 14 GPa the $B_u$ IR mode has imaginary 
frequency at $\Gamma$ point but also in other special and intermediate points of 
the BZ. In another way, the phonon spectrum does not present imaginary 
frequencies once In is eight-fold coordinated above 34 GPa. We see from 
Fig.~\ref{fig:7} (b) that in this phase of InVO$_4$ it is not depicted the phonon 
gap observed in the wolframite of InPO$_4$ and TiPO$_4$.~\cite{Lopez2012b} 

Group theory predicts that raspite and AgMnO$_4$-type structures 
have the  following vibrational representation for R and IR phonon modes at the 
$\Gamma$ point: $\Gamma$ = 18$B_g$ + 18$A_g$ + 16$B_u$ + 17$A_u$, i.e. 
36 R and 33 IR active modes. As we can see from Tables~\ref{table:5} and 
\ref{table:6} and from Fig.~\ref{fig:8} that all Raman and IR phonon modes shift 
to higher frequencies upon compression, i.e. the pressure coefficients and 
Gr\"uneisen parameters are positive. The phonon spectrum for these phases no 
present imaginary frequencies in all the range of pressure studied. For these phases 
there is not a gap in the phonon DOS like happens in other compounds with these 
structures such as CaSeO$_4$.~\cite{Lopez2015} In these phases there is not a 
clear separation of the internal and external modes due to the high coordination of In 
and V, also because their interatomic bond distances are different between each other, 
see the contributions of In and V to the phonon DOS of Fig.~\ref{fig:7}.

\subsection{Electronic properties}\label{sIII-D} 

Density functional theory has been used with success to describe the electronic 
structure of $AB$O$_4$ compounds.~\cite{Panchal2011,Ruiz2012,Lacomba2011} 
In particular, optical-absorption measurements in conjunction with \textit{ab initio}
calculations were used to study the electronic structure of zircon-type vanadates as 
function of pressure.~\cite{Panchal2011,Paszkowicz2015} However, it is well known 
that first principles calculations at the level of GGA underestimates the electronic 
band gap.~\cite{Zurek2015} Hence, some other approximations can be used in order 
to get a better estimation of the electronic band gap value, like 
GGA+\textit{U},~\cite{Lopez2009b} \textit{GW}~\cite{Shishkin2007} or using 
hybrid functionals like HSE03 or HSE06.~\cite{Pair2006,Hermann2016} Nevertheless, 
calculations with \textit{GW} or with hybrid functionals require much more computer 
time than the calculations with GGA or GGA+\textit{U}. So, their use in the community 
is still limited.

\begin{figure*}[t!]
\centering
\begin{tabular}{cccc}
\includegraphics[height=6.cm]{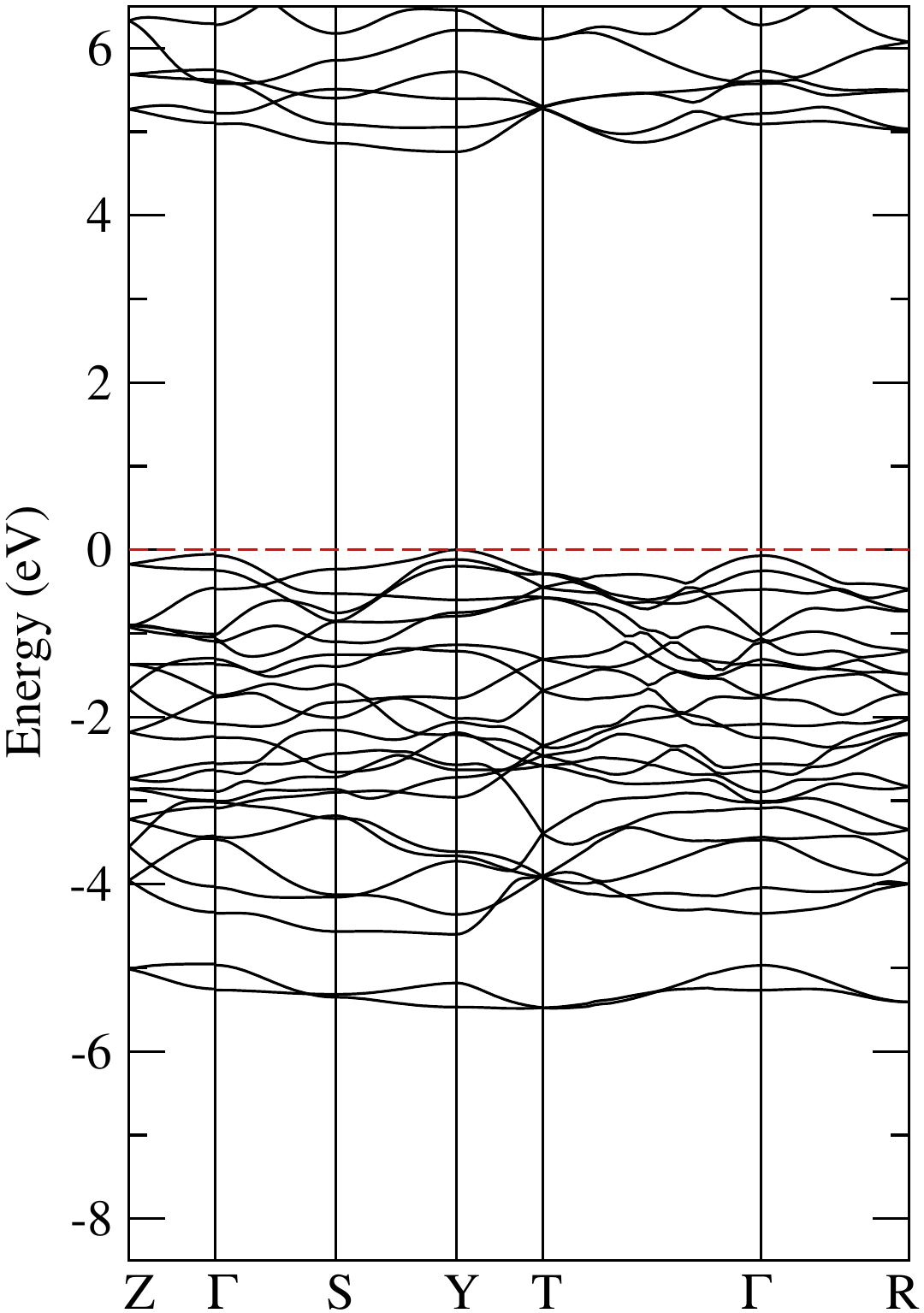} &
\includegraphics[height=6.cm]{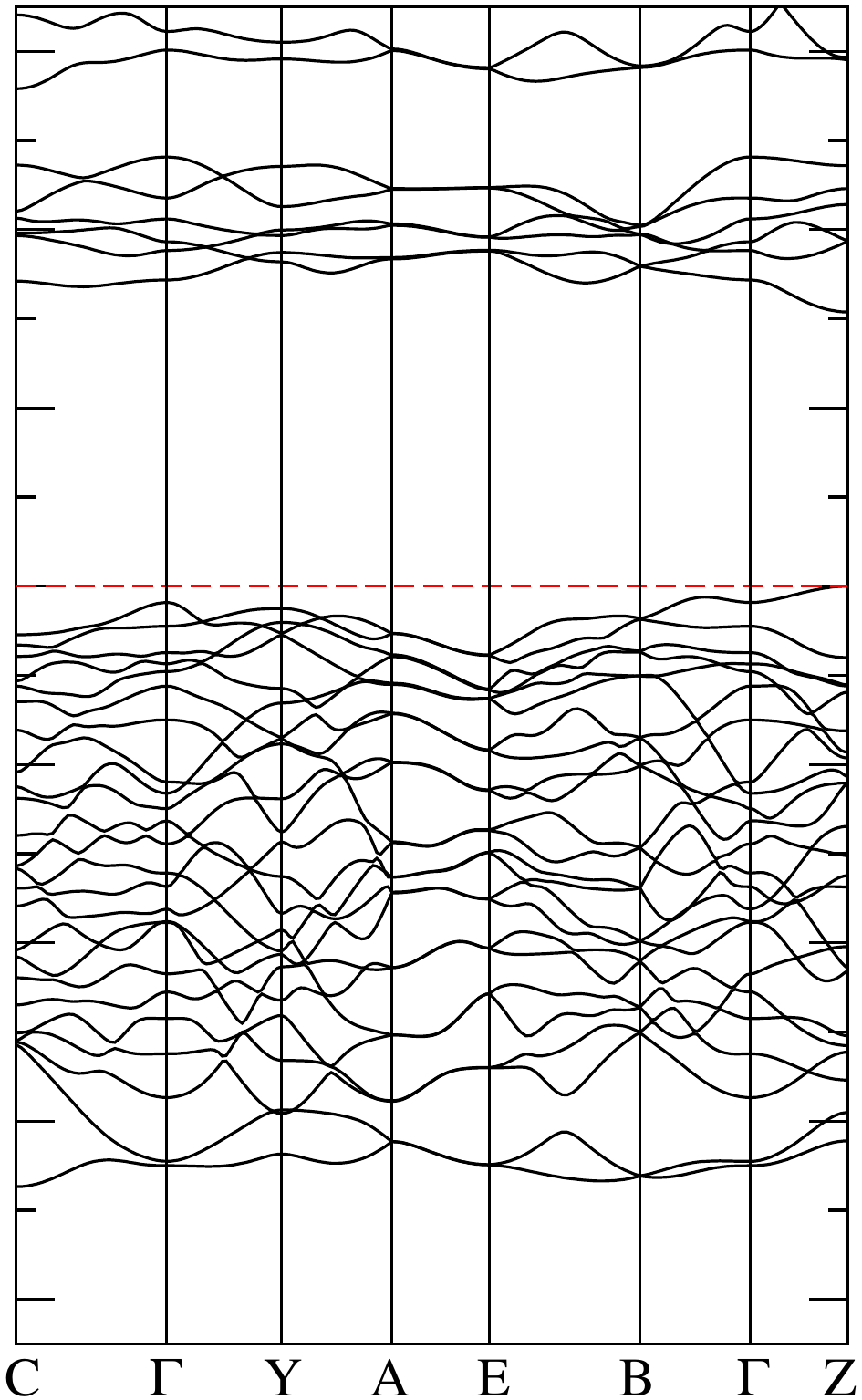} &
\includegraphics[height=6.cm]{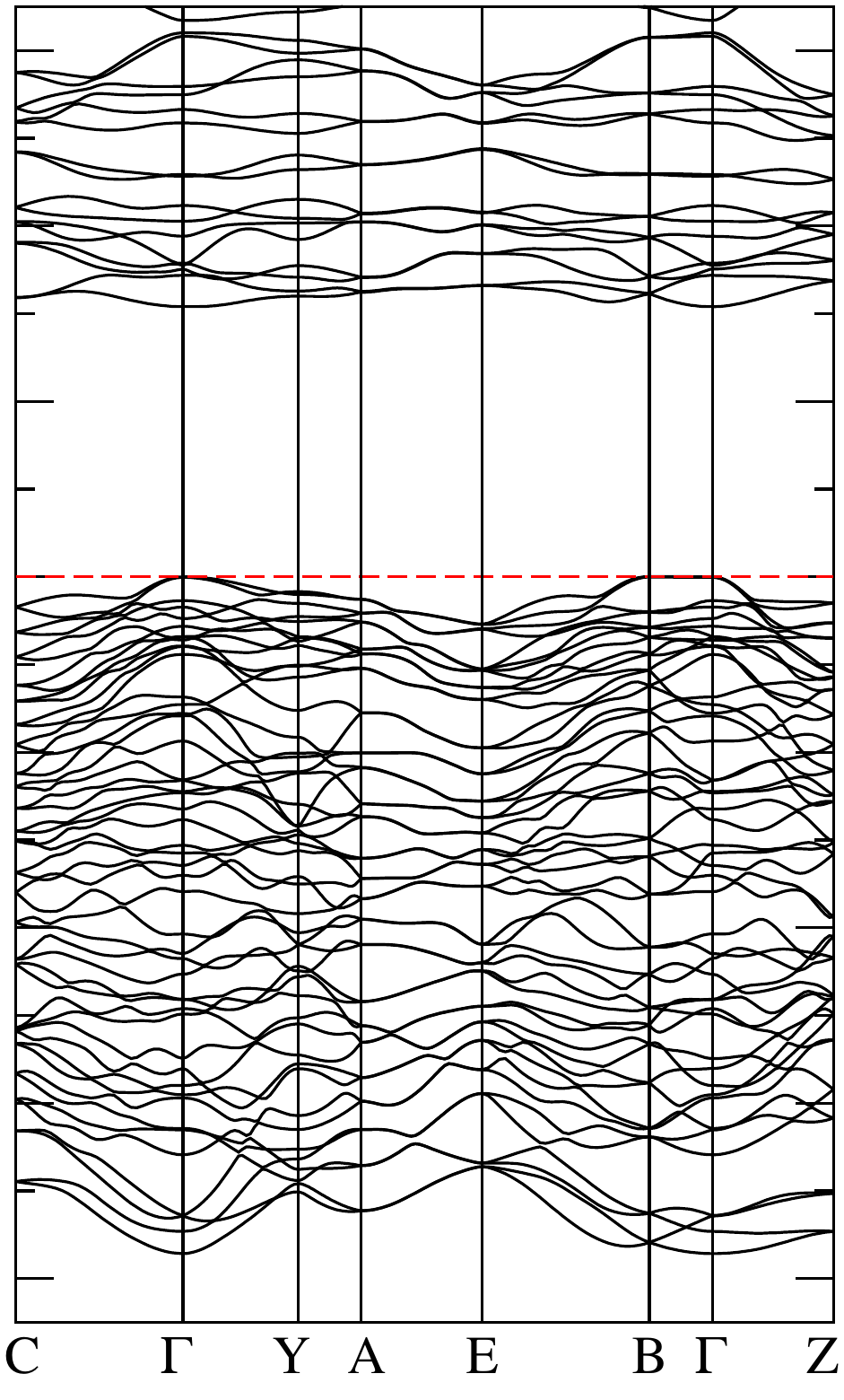} &
\includegraphics[height=6.cm]{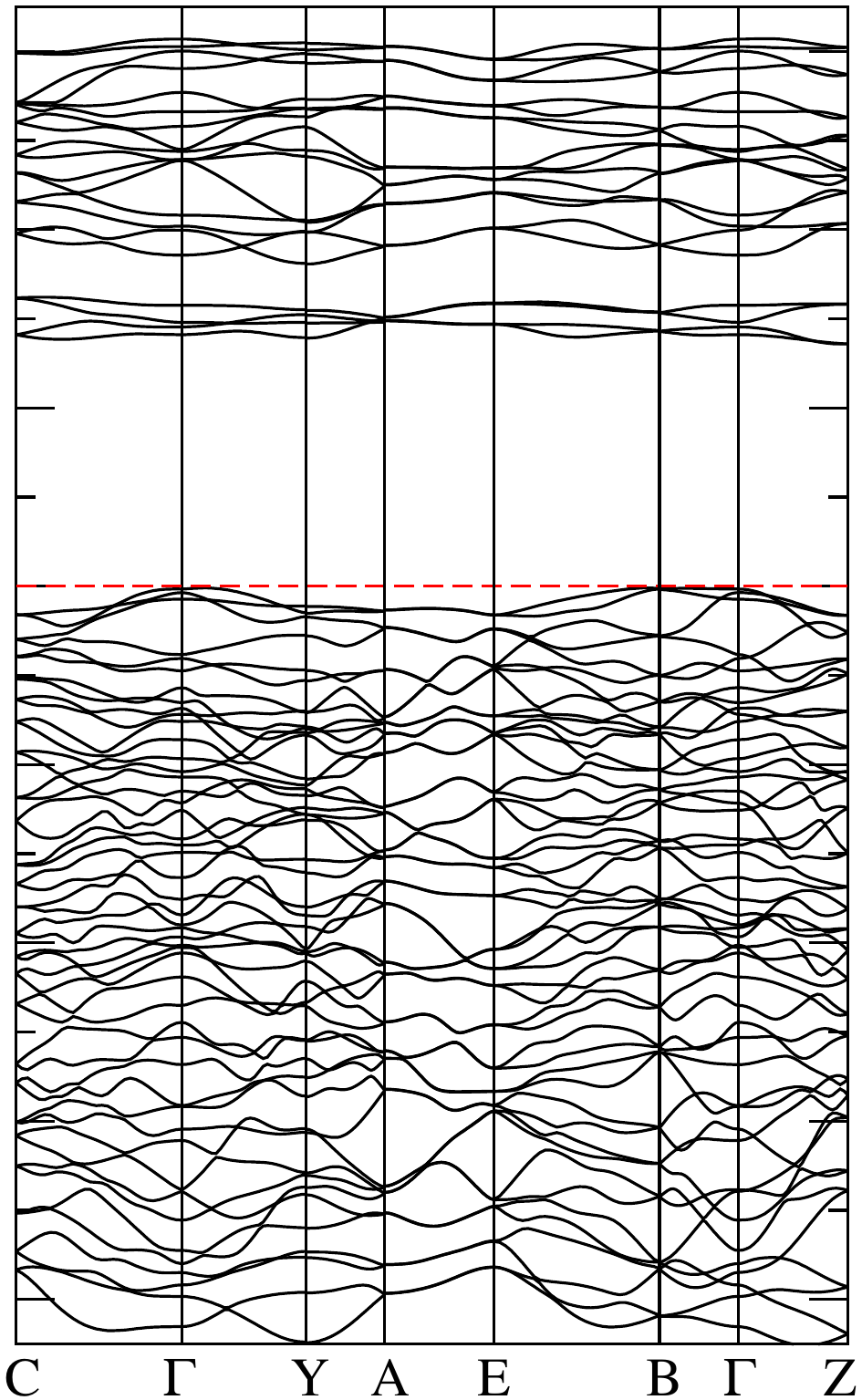} \\
\includegraphics[height=4.cm]{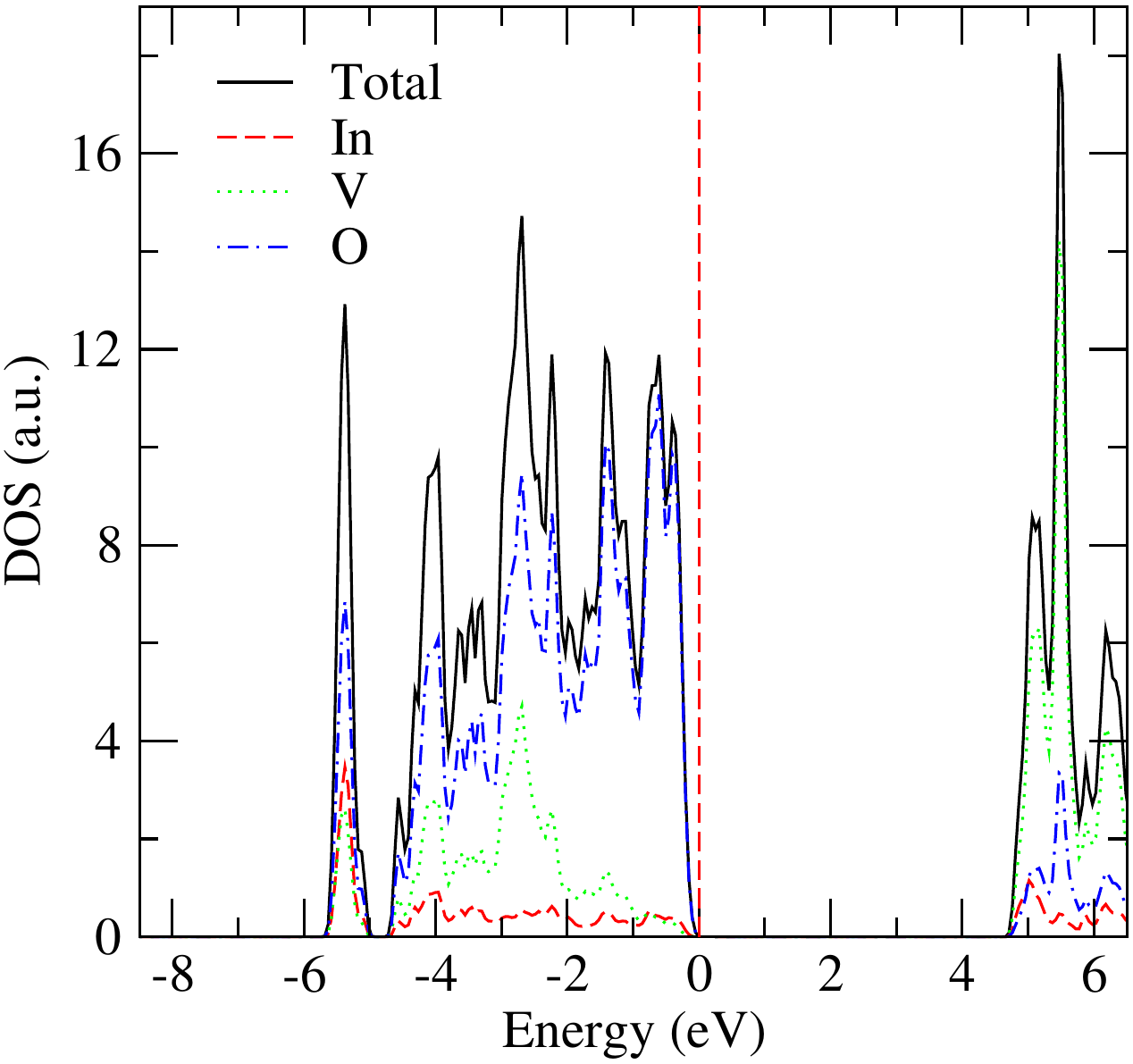} &
\includegraphics[height=4.cm]{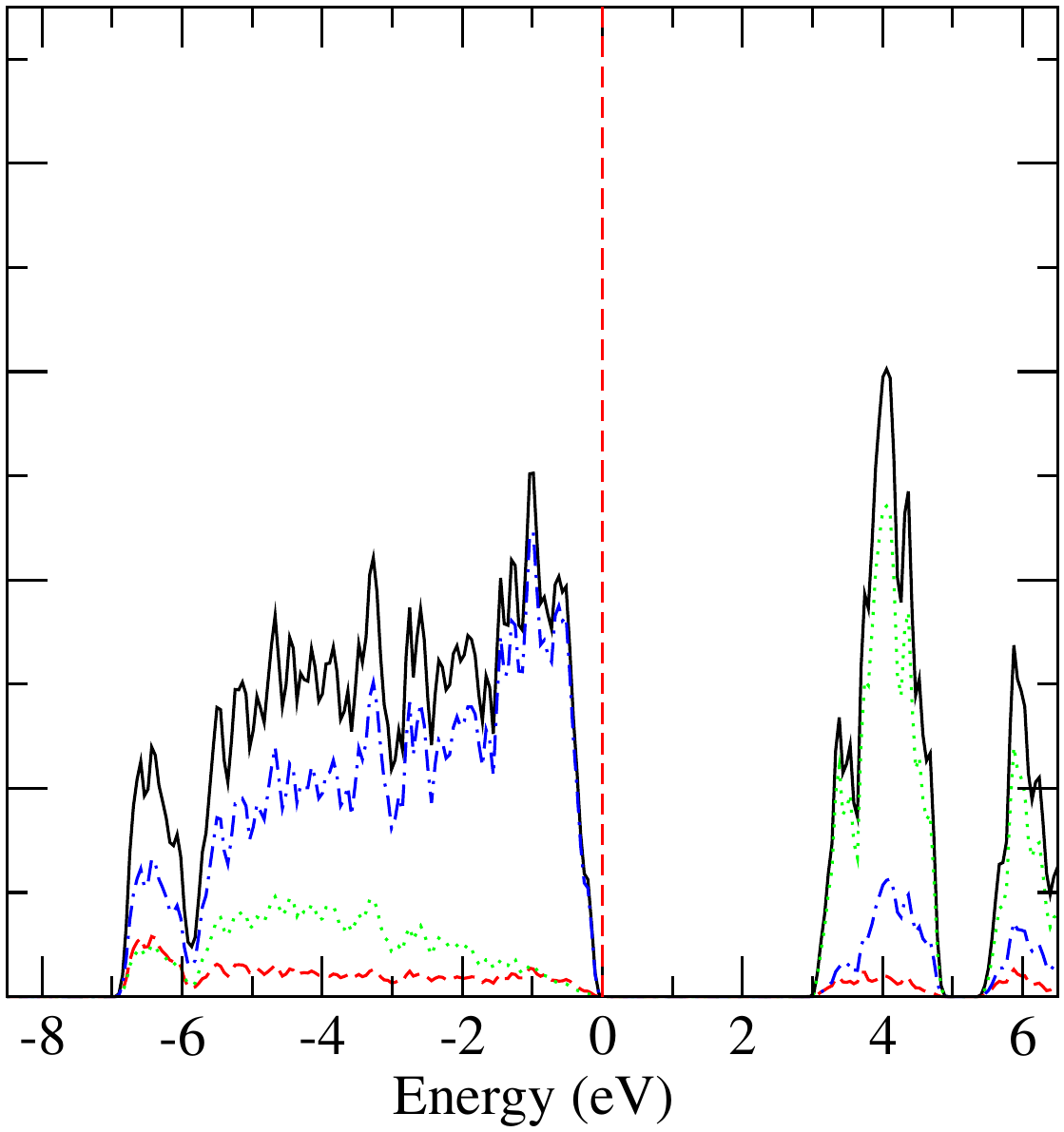} &
\includegraphics[height=4.cm]{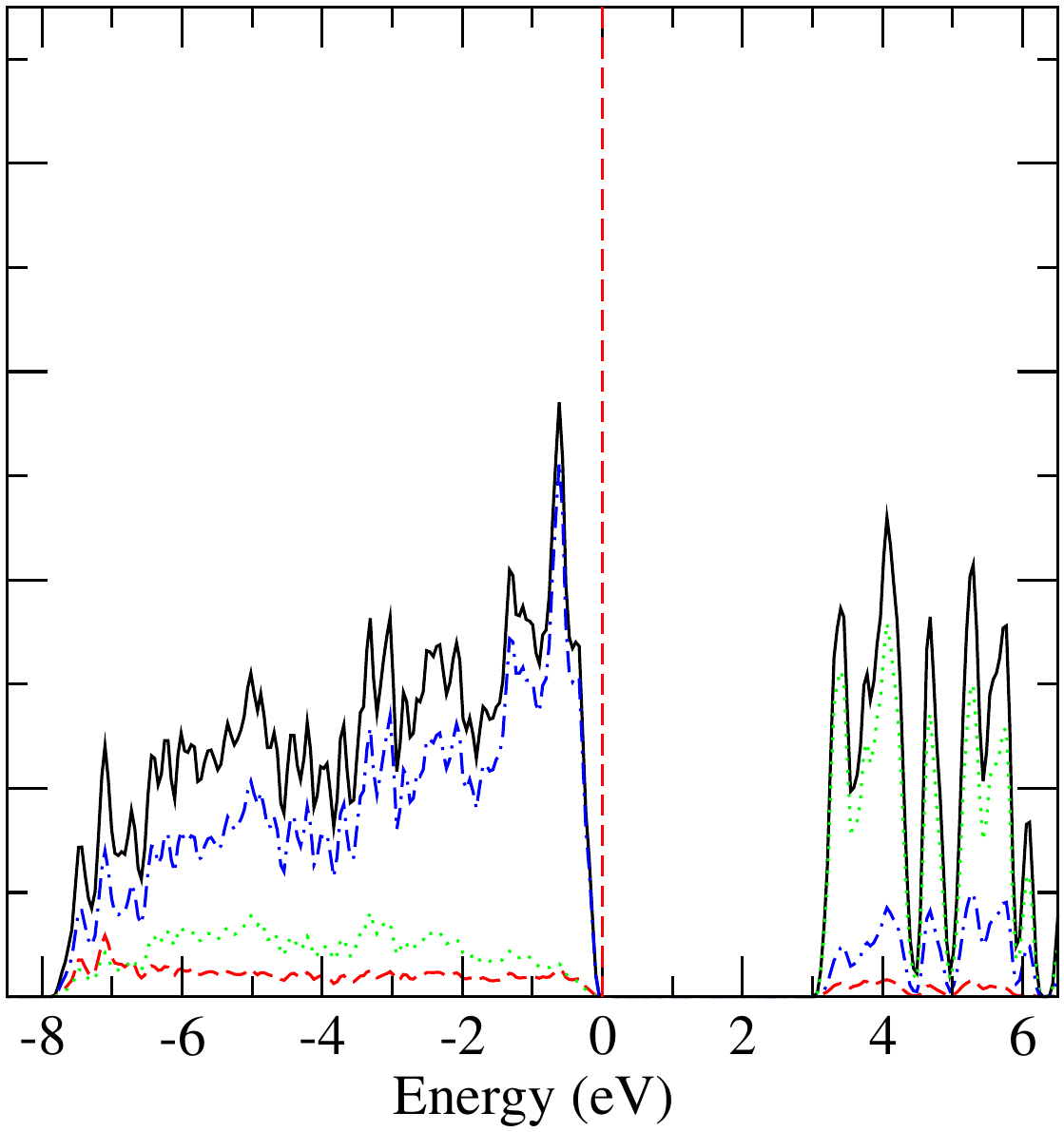} &
\includegraphics[height=4.cm]{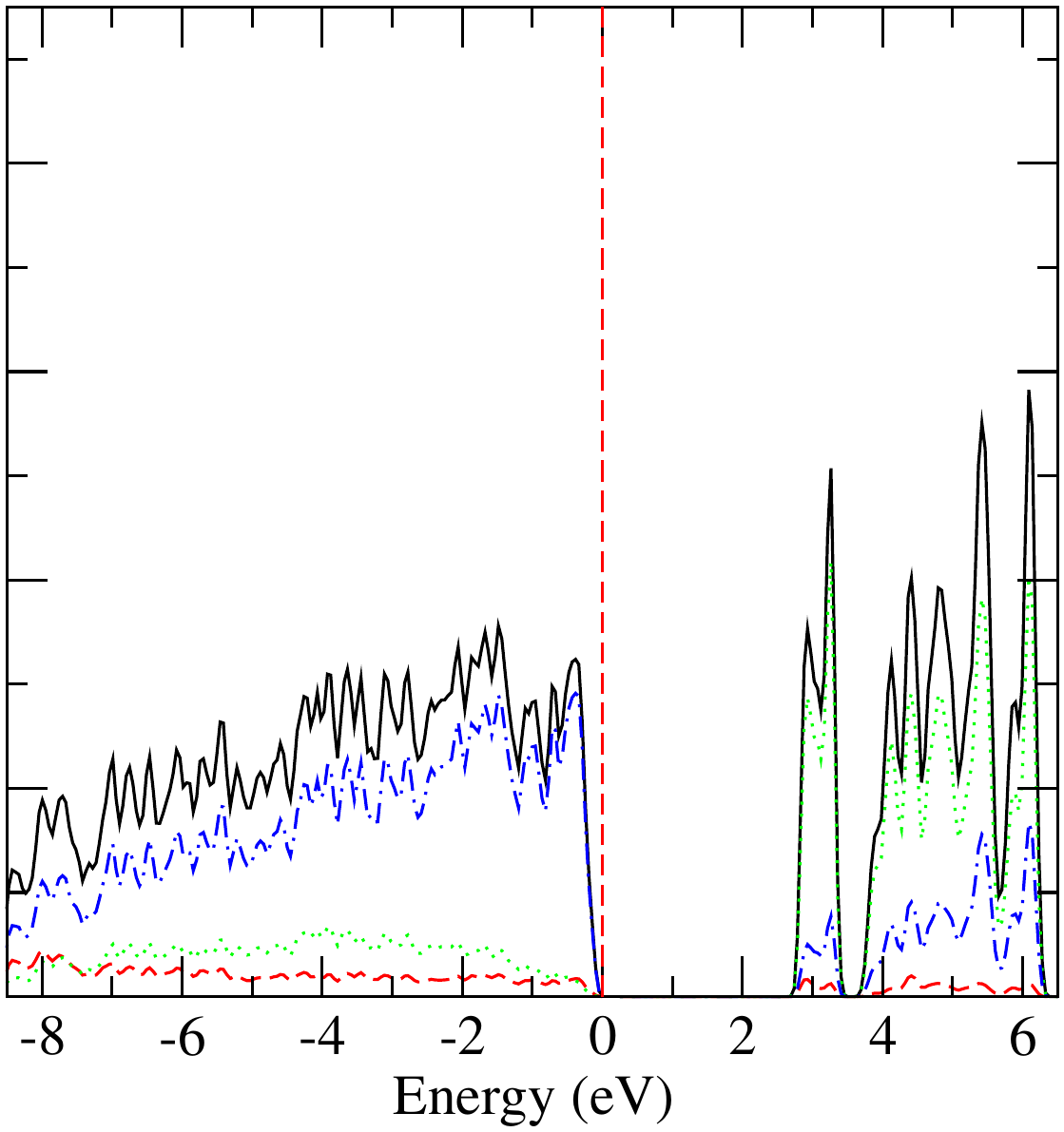} \\
(a) CrVO$_4$-type & (b) wolframite & (c) raspite & (d) AgMnO$_4$-type \\
at $\approx$ 0 GPa & at 8 GPa & at 30 GPa & at 47.7 GPa
\end{tabular}
\caption{(Color online) Band structure and partial density of states (DOS) of the 
most stable polymorphs of InVO$_4$: (a) CrVO$_4$-type (at $\approx$ 0 GPa), 
(b) wolframite (at 8 GPa), (c) raspite (at 30 GPa), and (d) AgMnO$_4$-type (at 
47.7 GPa).} 
\label{fig:9}
\end{figure*}

According to the literature, an experimental electronic band gap value of 3.2 
eV was reported for a thin-film of InVO$_4$-III.~\cite{Enache2009} While, 
theoretical calculations of bulk InVO$_4$-III preformed with WIEN2k code 
using PBE~\cite{Li2011} and Tran-Blaha Modified Becke-Johnson 
(TB-mBJ)~\cite{Mondal2016} exchange correlation functionals, reported a 
direct Y-Y electronic band gap of  3.24 and 4.02 eV, respectively. On the other 
hand, Li \textit{et al.}~\cite{Li2011} took the experimental data from 
Ref.~\citenum{Enache2009} and use the Lambert-Beer's law~\cite{He1996} 
to determine that the experimental electronic band gap of InVO$_4$ should be 
3.8 eV.  These results would comply with the fact that the electronic gap that is 
obtained at the GGA level is always lower than that observed experimentally. 
Given the observed differences in the value of the electronic band-gap, we 
consider that this compound should be studied again both experimentally as 
well as theoretically.

\begin{figure}[t!]
\centering
\begin{tabular}{c}
\includegraphics[width=8.5cm]{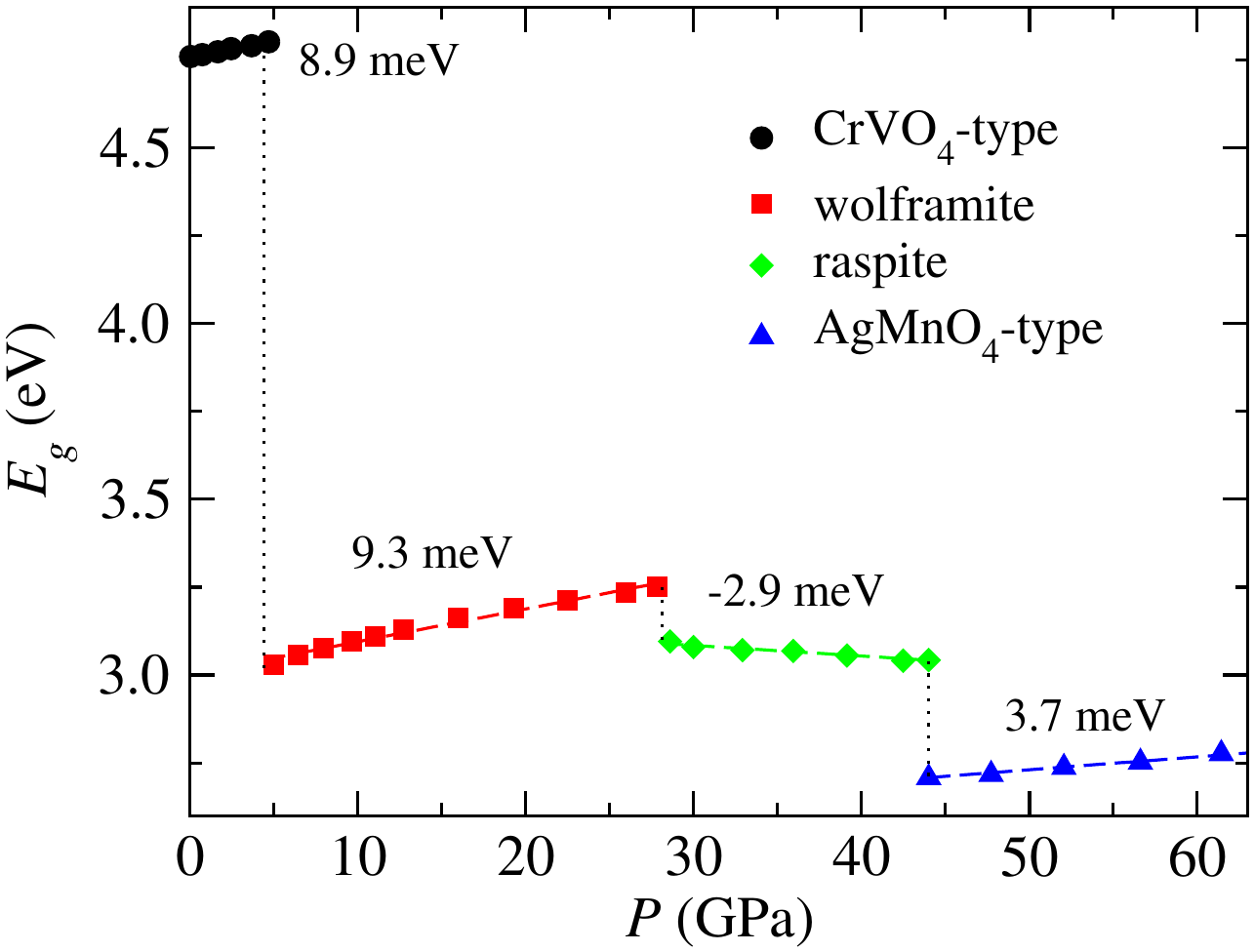} \\
\end{tabular}
\caption{(Color online) Pressure evolution of gap energy 
of InVO$_4$.} 
\label{fig:10}
\end{figure}

In this section we describe the results about the electronic structure of the most 
stable polymorphs of InVO$_4$ and the pressure evolution of the energy gap. 
In order to obtain a better description of electronic structure, we performed the
calculations by using the HSE06 hybrid functional. The details about the 
optimization of the crystal structure with this functional are described in 
Section~\ref{sII}. Figure~\ref{fig:9} shows the band structure and the partial 
density of states for (a) the CrVO$_4$-type, (b) wolframite, (c) raspite, and (d) 
AgMnO$_4$-type phases at the respective pressure. Whereas Fig.~\ref{fig:10} 
shows the pressure evolution of the energy band gap of InVO$_4$.

According to our calculations of band structure, Fig.~\ref{fig:9} (a), the phase III of 
InVO$_4$ is a direct band-gap material (top of valence band at Y and bottom of 
conduction band at Y). Where the top of the valence band consists mainly of
O 2$p$ states. The bottom of the conduction band is dominated by V 3$d$ states 
with a not negligible contribution of O 2$p$ and In 5$s$ states. Similar results were
obtained in Refs.~\citenum{Mondal2016} and \citenum{Li2011}. For this phase a 
direct electronic band-gap of 4.76 eV was obtained. On the other hand, we found that 
InVO$_4$-III presents an indirect electronic band-gap (top of valence band at 
$\Gamma$ and bottom of conduction band at Y) of 3.06 eV when the AM05 
functional is used.

As pressure is applied the band-gap in the CrVO$_4$-type phase increases with a 
pressure coefficient of 8.9 meV/GPa, as seen on Fig.~\ref{fig:10}. At this point we 
have to mention that there are important differences with the results obtained with 
AM05 functional. Where, we observed that as pressure increases the material changes 
from an indirect to a direct ban-gap at Y point with a pressure coefficient of -4.2 eV. 

At the phase transition from CrVO$_4$-type structure to wolframite we observed 
a collapse of the band gap which becomes close to 1.77 eV. In this phase we observed 
that the top of the valence band and the bottom of conduction band are located at 
$Z$ point. Again, the orbital contribution to the top of the valence band and the bottom 
of the conduction band is similar than in the low-pressure phase but the conduction 
band are also populated with small contributions from O 2$s$ and In 5$p$. The 
band gap collapse is consistent with the observation made in the study performed 
by Errandonea \textit{et al.} in  Ref.\citenum{Errandonea2013}. 
On the other hand, the theoretical results from Ref.~\citenum{Mondal2016} show that 
wolframite phase is an indirect band-gap material with the top of the valence band 
at $\Gamma$ and the bottom of the conduction band at $Z$. Our findings show that,
as in the band-gap behavior of CrVO$_4$-type structure under pressure, the band gap 
of wolframite increases with pressure with a pressure coefficient of 9.3 meV/GPa.
Almost the same value was obtained with the AM05 functional. Similar values 
were observed in the high pressure studies of some wolframates such as MgWO$_4$, 
ZnWO$_4$ and CdWO$_4$.~\cite{Ruiz2012}  

As seen on Fig.~\ref{fig:9} (c) raspite phase has a direct band-gap at $\Gamma$ 
point. While AgMnO$_4$-type behave like a indirect band-gap material with the 
top of the valence band at $\Gamma$ and the bottom of the conduction band at $Z$ 
point, Fig.~\ref{fig:9} (d) . In these phases the top of the valence band is almost all 
populated by O 2$p$ states, whereas the bottom of the conduction band is mainly 
occupied by V 3$d$ states. In these phases are produced additional reductions of the 
band-gap value, becoming the band-gap of AgMnO$_4$-type phase 2.8 (1.2) eV at 
60 GPa with the HSE06 (AM05) functional. Consequently, the band gap of InVO$_4$ 
changes from 4.76 (3.06) eV at ambient pressure to 2.8 (1.2) eV at 60 GPa with the 
hybrid HSE06 (AM05) functional. Such a large change of the electronic band gap 
has been observed in PbCrO$_4$ for $AB$O$_4$ oxides.~\cite{Errandonea2014} 

In this study we observe significative differences when the electronic structure 
is calculated with the GGA AM05 and the hybrid HSE06 exchange correlation 
functional, being the most significative differences on the electronic band-gap 
values, the determination if the band-gap is direct or indirect, and the slope of the 
band-gap pressure coefficients. Unfortunately, there are no compelling experimental 
studies in the literature to support or disprove our findings. We hope that this work 
will serve to encourage the experimental scientists to study the electronic structure 
of InVO$_4$ at ambient conditions and under pressure.

\section{Summary and Conclusions}\label{sIV}

We presented a first principles study of structural, electronic and vibrational 
properties of InVO$_4$ from ambient pressure to 62 GPa. The quasi-harmonic 
approximation has been used to obtain the Gibbs free energy and determine the 
phase transitions at ambient temperature. Where a good agreement between our 
theoretical results and the reported experimental data was obtained. In our study 
we found that wolframite presents o drastic change in the interatomic bond 
distances in order to increase the coordination of In at elevated pressures, which 
has an important effect in the Raman and infrared phonon frequencies at 
$\Gamma$ point, but also in the branches of the phonon spectrum in other points 
of the Brillouin zone. It has been observed that the characteristic acoustic $B_u$ 
infrared phonon mode of wolframite, the phonon mode that has negative 
pressure coefficient and Gr\"uneisen parameter, softens completely around 14 
GPa. Which is related with the instability of the wolframite phase as pressure 
increases. Besides, in our study two new high pressure phases were observed 
above 28 GPa, being the raspite and AgMnO$_4$-type structure. As is known 
the last one was observed as post-scheelite phase in CaSeO$_4$ and as a
high-pressure phase of CaSO$_4$.

It has been proposed that pressure could induce the metallization of orthovanadates 
at relative low pressure (11 GPa)~\cite{Garg2013} however our calculations show 
that InVO$_4$ does not become metallic up to 60 GPa. 

The information of transition pressure and volume reduction involved in the phase 
transition sequence observed in InVO$_4$ in this study from phase III (CrVO$_4$-) to 
VII (AgMnO$_4$-type structure) can be summarized as follows: 
{\small{
\begin{eqnarray*}
{\mathrm {III}}\xrightarrow[\Delta V=-16.8\ \%]{\;\; P_T=4.4\ {\mathrm {GPa}} \;\;
}\, {\mathrm {V}} \xrightarrow[-6.5]{\;\; 28.1 \;\;}\, {\mathrm {VI}} \xrightarrow[-3.5]{\;\; 44 \;\; }\, {\mathrm {VII}}
\end{eqnarray*}}}\\
\\
{\bf Author Information} \\
{\bf Corresponding Author} \\
*E-mail: sinlopez@uacam.mx \\
{\bf Author Contributions} \\
The manuscript was written through contributions of all authors. All authors have 
given approval to the final version of the manuscript. \\
{\bf Notes} \\
The authors declare no competing financial interests.

\begin{acknowledgement}
This work has been done under partial financial support from Spanish MINECO 
under projects MAT2013-46649-C4-1/3-P, MAT2015-71070-REDC, and 
MAT2016-75586-C4-1/3-P. S.M.L. thanks CONACYT from Mexico for financial 
support through the program "C\'atedras para J\'ovenes Investigadores". We thank 
the computer time provided by the RES (Red Espa\~{n}ola de Supercomputaci\'on) 
and the MALTA cluster.
\end{acknowledgement}

\pagebreak


\providecommand{\latin}[1]{#1}
\providecommand*\mcitethebibliography{\thebibliography}
\csname @ifundefined\endcsname{endmcitethebibliography}
  {\let\endmcitethebibliography\endthebibliography}{}

\end{document}